\documentclass[oldversion]{aa}
\usepackage{graphics}
\usepackage{natbib}
\usepackage{txfonts}
\usepackage{multirow}
\input{psfig}
\bibpunct[; ]{(}{)}{,}{a}{}{,}
\begin{document}

%   \thesaurus{06     % A&A Section 6: Form. struct. and evolut. of stars
%              (03.11.1;  % Cosmogony,
%               16.06.1;  % Planets and satellites: general,
%               19.06.1;  % Solar system: general,
%               19.37.1;  % Stars: formation of,
%               19.53.1;  % Stars: oscillations of,
%               19.63.1)} % Stars: structure of.
%
   \title{Young star clusters in interacting galaxies - NGC 1487 and 
          NGC 4038/4039\thanks{Based on observations collected
	    at the European Southern Observatory, Chile, programme identification 
	    numbers 63.N-0528, 65.N-0577, 66.B-0419, 67.B-0504 and 68.B-0530}}

\author{Sabine Mengel$^1$, Matthew D. Lehnert$^2$, Niranjan A. Thatte$^3$, William D. Vacca$^4$, Brad Whitmore$^5$ and Rupali Chandar$^6$}

\authorrunning{Mengel et al.}

\offprints{S. Mengel}

\institute{European Southern Observatory, Karl-Schwarzschild-Str. 2, D-85748 Garching, Germany\\
        \and
           Observatoire de Paris, CNRS, Universite Denis Diderot; 5, Place Jules Janssen, 92190, Meudon, France\\
        \and
	   University of Oxford, Dept. of Astrophysics,
	   Denys Wilkinson Building, Keble Road, GB-Oxford OX1 3RH\\
	\and
	   Stratospheric Observatory for Infrared Astronomy/Universities Space Research Association, 
	   NASA Ames Research Center, Moffett Field, CA 94035, USA\\
	\and
	   Space Telescope Science Institute, 3700 San Martin Drive, Baltimore, Maryland 21218, USA\\
        \and
           The University of Toledo, Toledo, OH 43606, USA\\
           email: smengel@eso.org, matthew.lehnert@obspm.fr, thatte@astro.ox.ac.uk, wvacca@sofia.usra.edu, 
                  whitmore@stsci.edu, Rupali.Chandar@utoledo.edu}

\date{Received ............... ; accepted ............... }

\abstract
{We estimate the dynamical masses of several young ($\approx10$~Myr) 
massive star clusters in two interacting galaxies, NGC~4038/4039 
("The Antennae") and NGC~1487, under the assumption 
of virial equilibrium.  
These are compared with photometric mass estimates from $K$-band 
photometry and assuming a standard Kroupa IMF.
The clusters were selected to have near-infrared colors dominated 
by red supergiants, and hence to be old
enough to have survived the earliest phases of cluster evolution when
the interstellar medium is rapidly swept out from the 
cluster, supported by the fact that there is no obvious H$\alpha$ emission
associated with the clusters.
All but one of the Antennae clusters have dynamical and photometric mass
estimates which are within a factor $\approx2$ of one another, implying both that 
standard IMFs provide a good approximation to the IMF of these 
clusters, and that there is no significant extra-virial motion, 
as would be expected if they were rapidly dispersing.  
These results suggest that almost all of the Antennae clusters in our
sample have survived the gas removal phase as bound or marginally bound
objects. 
Two of the three NGC~1487 clusters studied here have 
M$_{dyn}$ estimates which are significantly larger than the
photometric mass estimates.
At least one of these two clusters, and one
in the Antennae, may be actively in the process of dissolving.  The
process of dissolution contributes a component of non-virial motion to
the integrated velocity measurements,
resulting in an estimated  M$_{dyn}$ which is too high relative to
the amount of measured stellar light. 
The dissolution candidates in both galaxies are amongst the
clusters with the lowest pressures/densities measured in our sample.
}


   \maketitle

%
%________________________________________________________________

\section{Introduction}
Despite being the targets of intensive studies over the last fifteen years,
young extragalactic star clusters, which are found in large numbers in
interacting galaxies
\citep[e.g.][]{Hetal92,Wetal93,W99,Zepf99,Mengel05,Bastian06,Trancho07}, 
as well as in other
environments like normal spirals \citep{LarsenRichtler04, Larsenetal04}, seem 
to have raised more questions than they have answered.

One of the most obvious, but arguably most interesting, questions is: how many
of the young star clusters (YSCs) survive to old age (i.e. become globular clusters),
and what happens to the others? Most likely many clusters disperse, contributing to
the general field star population; although, it remains uncertain what fraction
of the general field population originated in stellar clusters.

Several studies \citep[e.g.][]{Larsenetal04,Bastian06} have shown that the properties of 
(at least some) young clusters
are consistent with them being the progenitors of what we see
as globular clusters today. Is it possible to identify from a population
of extragalactic young star clusters those which will survive
for a Hubble time? Or, as a different way to phrase the same problem:
How many star clusters, and with which properties, formed the host
population of the globular clusters in today's galaxies?

In environments as different as mergers like NGC 4038/4039 and the Milky Way,
it seems that, at least up to around 100 Myr, 50-90\% of the star clusters
are destroyed within each decade of time. This effect has been named ``infant mortality''.

The current hypothesis \citep{Hills80, Lada84, BoilyKroupa03a, BoilyKroupa03b, Fall05, 
GoodwinBastian06, Whitmore07} is that the gas removal caused by stellar
winds and supernovae unbinds some of the clusters, and that this process
is dominant only out to roughly 30 Myr. Later, the much slower and less destructive
two-body relaxation takes over, which has, together with other effects like
the impact of the galactic gravitational field etc., the potential to dissolve
another fraction of the initial survivors.

More observational data are necessary to get a clearer idea of the
dynamical processes at work during cluster formation and destruction.
The cluster populations analysed so far with respect to their 
ages have not been corrected for the (unknown) cluster formation history.
However, all studies which analyse statistically significant
numbers of high-mass clusters (NGC 4038/4039, \citet{Fall05, Mengel05}, 
M51, \citet{Bastian05}) are interacting systems where the star/cluster
formation history is neither constant, nor a delta burst, but rather
some more complex, unknown funtion of time. This certainly affects
the age distribution of clusters and hence the destruction rate
derived from it. 

A different approach targets individual star clusters for intense studies
of their physical parameters, with the goal of using these paramters
to decide whether a star cluster is doomed or a candidate for a future GC.

In our original study \citep{Mengel02} we assumed (as had, for example,
\citet{HF96a, HF96, Sternberg98}) that clusters are in virial equilibrium,
since at ages of around 8 Myr, they have survived for many crossing
times. However, in the view of the high cluster destruction rate derived from
recent studies, this assumption may not be universally applicable.

Our current study expands the number of analysed individual clusters.
With a larger sample, we hope to be able to find a diagnostic to
determine the dynamical state of an extragalactic, and hence only
barely resolved, star cluster. Or at least to see which potential
techniques are unfeasible or do not lead to useful results.

The galaxies we targeted are NGC 4038/4039 and NGC 1487. While the first,
also called ``The Antennae'', is one of the best studied nearby mergers,
NGC 1487 is less well known. It is described as a peculiar galaxy, which shows
two faint tails as a tracer of earlier interaction. \citet{LeeLee} conclude
from their two-colour analysis of the cluster system that the merging
process could have taken place 500 Myr ago. Most of the star clusters
are found in three or four ``condensations'', and the brightest clusters,
like those targeted for our study, are much bluer than the larger population
of fainter clusters. In total, \citet{LeeLee} found more than 500
cluster candidates in HST/WFPC2 data. 
The galaxy is at approximately half the distance to
the Antennae, but substantially fainter: Its total magnitude is comparable
to the LMC.

The stellar velocity dispersion in the clusters is typically
around 15 km s$^{-1}$ and therefore requires medium- to high spectral resolution of
these faint targets, which is only achievable with 10m class telescopes.
Apart from near-infrared imaging for the cluster photometry, we 
need an estimate of the cluster
size, which for objects at distances between 10 and 20 Mpc and sizes of
2-4 pc requires very high spatial resolution.

\section{Observations \& Data Reduction}\label{observations}

\begin{figure*}
\begin{center}
\psfig{figure=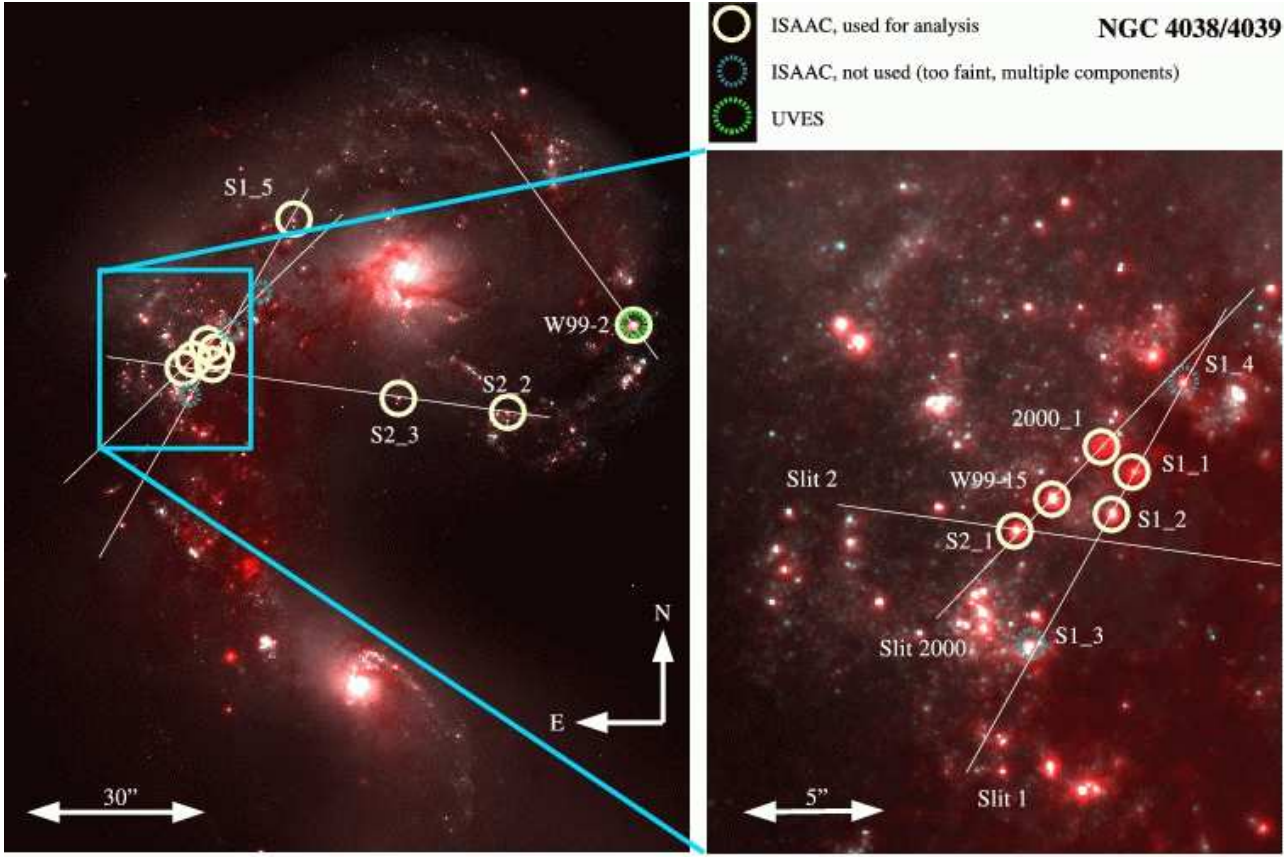,width=17cm,angle=270}
\caption{\label{antennaefig} False-colour image of NGC 4038/4039 
with HST/ACS F814W covering both the blue and the green channel,
and VLT/ISAAC Ks in the red channel. Clusters which are presented in this
publication are marked, using the naming convention as in W99 (those
which had been listed there), or according to slit number.}
\end{center}
\end{figure*}

\begin{figure*}
\begin{center}
\psfig{figure=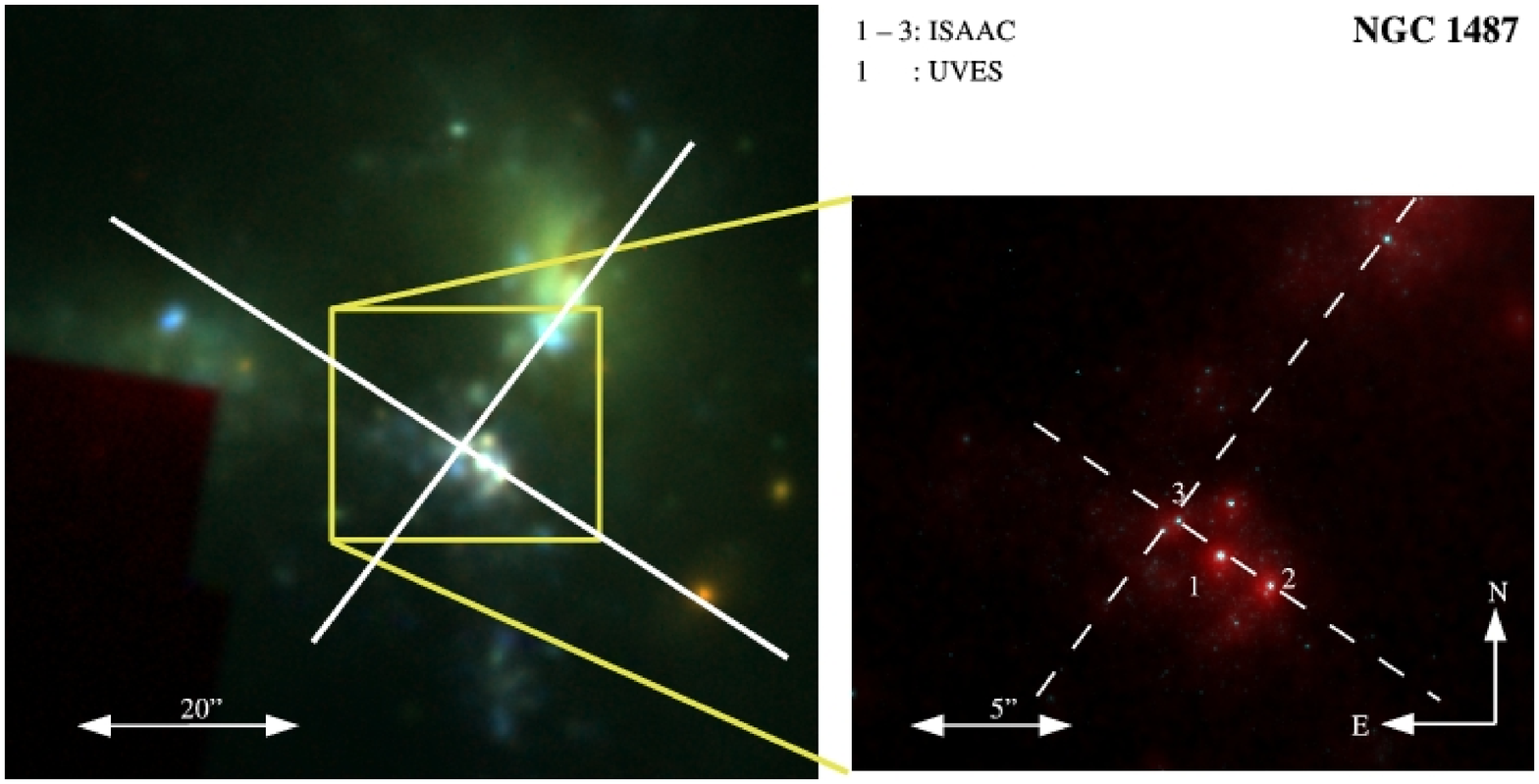,width=17cm}
\caption{\label{n14fig} Left:  NGC 1487 false colour image, with 
HST/WFPC2 F435W in the blue,HST/WFPC2 F814W in the green, and NTT/SOFI Ks in the red
channel. HST images were smoothed to the FWHM of the SOFI images.
Right: False-colour image with the SOFI Ks image in the red, and  HST ACS/HRC F814W 
(not smoothed, therefore substantially better spatial resolution) in green/blue.
The clusters for which we present spectra are labelled in the image on the right:
Clusters 1-3 were observed with ISAAC, cluster 1 additionally with UVES.
During the largest fraction of the observations, the slit was oriented such that
it covered all three clusters. Cluster 3 was covered during another integration
which accounts for roughly 20\% of its exposure time (see Tab.\ref{exptimestable}).}
\end{center}
\end{figure*} 

%\section{Reduction and  Analysis}

In this section we present the ground based imaging and spectroscopic
data obtained for the K-band bright clusters with strong CO absorption 
(as a consequence, the ages span a narrow range around 8.7 Myr) selected
for spectroscopy.  Table \ref{exptimestable} lists
integration times for both imaging and spectroscopy.  Supporting
archival images taken with the {\em Hubble Space Telescope} (HST) are also
described.

\begin{table*}
\begin{center}
\caption{Integration times of all images and spectra used for the main analysis of the
clusters. We did not list the many different narrow band filters which were used for
age dating (Br$\gamma$, CO2.32, Pa$\beta$, H$\alpha$). 
\label{exptimestable} }
\begin{tabular}{l|cc|cccc}
\hline
\hline
Cluster     &   \multicolumn{2}{c|}{Integration times VLT/NTT} & \multicolumn{4}{c}{Integration time HST/ACS} \\
             & ISAAC spectroscopy & ISAAC/SOFI Ks imaging & F814W & F550M & F435W & F555W\\\hline
$[$W99$]$2    &  2400s    &  360s   &     3360s     &  2544s         &  -     &  1530s \\
$[$W99$]$15   & 16800s    &  360s   &     3360s     &  2544s         &  -     &    -   \\
S1\_1         & 28800s    &  360s   &     3360s     &  2544s         &  -     &    -   \\
S1\_2         & 28800s    &  360s   &     3360s     &  2544s         &  -     &    -   \\
S1\_3         & 28800s    &  360s   &     3360s     &  2544s         &  -     &    -   \\
S1\_4         & 28800s    &  360s   &     3360s     &  2544s         &  -     &    -   \\
2000\_1       & 16800s    &  360s   &     3360s     &  2544s         &  -     &    -   \\
S1\_5         & 9000s     &  360s   &     3360s     &  2544s         &  -     &    -   \\
S2\_1         & 9000s     &  360s   &     3360s     &  2544s         &  -     &    -   \\
S2\_2         & 9000s     &  360s   &     3360s     &  2544s         &  -     &    -   \\
S2\_3         & 9000s     &  360s   &     3360s     &  2544s         &  -     &    -   \\\hline
NGC 1487-1    & 14700s    &  100s   &     640s      &    -           & 1540s  &    -   \\
NGC 1487-2    & 14700s    &  100s   &     640s      &    -           & 1540s  &    -   \\
NGC 1487-3    & 18300s    &  100s   &     640s      &    -           & 1540s  &    -   \\\hline\hline
\end{tabular}
\end{center}
\end{table*}

\subsection{SOFI and ISAAC imaging data}\label{IRimagingredandanal}

ISAAC/VLT imaging of NGC 4038/4039 was performed in ON/OFF mode during
the nights 15.04.2001 (Ks-band) and 16.04.2001 (CO-bandhead filter).
The target fit completely onto the detector (0\farcs1484/pixel, total
field size 2\farcm5 $\times$ 2\farcm5).  Seeing was excellent during both
of these photometric nights (the FWHM of the PSF from coadded frames is
$<$0\farcs4).
SOFI/NTT imaging of NGC 1487  covered J, H, Ks broad- and 
Br$\gamma$, CO2.32$\mu$m, Pa$\beta$, and continuum NB2.28$\mu$m, 
NB1.215$\mu$m and NB2.195$\mu$m narrow bands. Here the field size was
roughly twice that of ISAAC (4\farcm9 $\times$ 4\farcm9, with a pixel
size of 0\farcs292). With the target spanning only a bit more than
2',  an efficient on-chip offset pattern could be used.

Reduction of the ISAAC and SOFI broad- and narrow band data 
was performed using the IRAF
package\footnote{IRAF is distributed by the National Optical Astronomy
Observatories, which are operated by the Association of Universities
for Research in Astronomy, Inc., under cooperative agreement with the
National Science Foundation.}.  It included dark and sky subtraction
(either using the median of several neighbouring sky images or, where
this led to residuals, doing pairwise subtraction), and flat fielding by
a normalized median of all sky frames. All of the ON frames were slightly
offset with respect to each other, in order to minimize the effect of
pixel defects. Therefore, they had to be shifted to a common location
before using the {\sl imcombine} task (setting the minmax rejection
algorithm to reject the highest and the lowest pixel) to combine the
single frames. 

Photometric standards GSPC S279-F, Ks-magnitude 12.03 (ISAAC) and 
S301-D, Ks-magnitude 11.79 (SOFI) from \citet{Persson98}
were used for flux calibration of all the broad-band data.
The resulting zeropoints in Ks (the only ones we will give here,
because Ks photometry will be used in the cluster analysis) 
were 24.28 mag and 22.27 mag, respectively).

The target clusters were selected to have a high 
CO(3-1) band-head absorption equivalent width (which is covered by the ISAAC and SOFI
NB2.34$\mu$m filter), which revealed clusters at
ages $\approx10$~Myr, where the near-infrared emission is
dominated by red supergiants.  Clusters at ages which are dominated by
very hot young stars do not show photospheric absorption features, or
their absorption lines are rotationally broadened, making measurement
of their {\sl stellar} velocity dispersion very difficult.  Locations
of the selected clusters are shown in Figs.~\ref{antennaefig} and
\ref{n14fig}.  They must be detected in at least the I-band with
$HST$, in order to measure their size (see \S\ref{sizeestsection}).

\subsection{Spectroscopic data}\label{IRredandanalsection}

Spectroscopy was performed with ISAAC at VLT-ANTU 
in 04/2000, 04/2001 and 12/2001. ISAAC was configured to have an 0\farcs3 wide
slit, and a central wavelength of 2.31$\mu$m with a total wavelength
coverage from 2.25 to 2.37$\mu$m.  This was sufficient to include
the $^{12}$CO (2-0) and (3-1) absorption bands at a spectral resolution
$\lambda/\Delta\lambda$$\sim$9000.  Observations of late-type
supergiant stars were taken so that they could be used as templates for
the determination of the velocity dispersions\footnote{This research has
made use of the SIMBAD database, operated at CDS, Strasbourg, France}.
Observations were performed by nodding along the slit and dithering the
source position from one exposure to the next.  B5V atmospheric calibrator 
stars were observed several times during the night.

The reduction of ISAAC spectroscopy data also made use of the 
IRAF data reduction package. It included dark subtraction and
flat-fielding (using a normalized flat-field created from internal flat
observations) on the two-dimensional array.  Sky subtraction was performed
by typically using the median of 3-4 frames taken at the other nodding
position, and pairwise if the first strategy left strong residuals.
This was followed by a rejection of cosmic ray hits and bad pixels.
The spectra were then corrected for tilt and slit curvature by tracing the
peak of the stellar spatial profile along the dispersion direction and
fitting a polynomial to the function of displacement versus wavelength.
Our suite of velocity template stars was used for this purpose.

Wavelength calibration was performed in a similar manner. Here, we used
a combination of arc discharge lamp observations and night sky lines for the 
identification of wavelengthts.

Single object spectra were combined by shifting-and-adding, including
a rejection of highest and lowest pixels.  

The object spectra were then extracted
from user defined apertures (usually the limits of the apertures were
placed at the point where the counts in the combined spectrum had dropped
to 1/10 of the peak value).  A linear fit to the background (below
$\approx$7\% of the peak intensity for all clusters) on both sides of
the object spectrum was subtracted.

For cluster NGC 1487-3, care was required in both the slit positioning
and the spectrum extraction due to the nearby faint companion cluster.
For one slit orientation, covering the largest fraction of
the exposure time, clusters 2 and 3 were lined up in the slit, which 
offset it just slightly from the brightest cluster 1 
(see Figs. \ref{n14fig} and \ref{N1487_acq}). 
For 3600s, we integrated in an almost perpendicular slit orientation which included 
only cluster 3. Despite trying to avoid flux from the fainter companion
from entering the slit or the extraction aperture, and excellent seeing
during most of the exposure time, it is likely that some unknown 
contribution from the companion is present in our final spectrum for
cluster NGC 1487-3 (see also Sect. \ref{multiples}).

An atmospheric calibrator (B5V) was observed and reduced in the
same way as the target and used to divide out the atmospheric absorption
features from the spectra.

This article concentrates on ISAAC spectroscopy, and we are using only
the UVES velocity dispersions previously published \citep{Mengel02,Mengel03} for
clusters where both, ISAAC and UVES spectroscopy was perfomed:
[W99]-2 and NGC1487-1. 
Data reduction and analysis of the UVES spectra is described in those 
publications.

\subsection{Hubble Space Telescope Imaging  data}\label{HSTredandanalsection}

We use imaging observations taken with $HST$ to estimate the
size of each cluster.
Observations for the size determination of the NGC 1487 clusters were
obtained as part of Hubble Space Telescope Cycle 11 observations
(Proposal-ID 9473, PI: Vacca). We are using the F814W and F435W
Advanced Camera for Surveys / High Resolution Channel (ACS/HRC) images.
For the Antennae clusters we use HST/ACS-WFC images 
obtained for Proposal-ID 10188 (PI: Whitmore) in the F550M and F814W
filters.  In addition, we took advantage of higher resolution
ACS/HRC images in the F555W filter of a supernova in the Antennae
(Proposal ID 10187, PI: Smartt), which happened to include one of our 
target clusters ([W99]-2).

For all $HST$ data, we use the pipeline reduced images.  Total
integration times are listed in Tab \ref{exptimestable}.  
The total field coverage for the Antennae WFC F814W and F550M images
was roughly 3\farcm5x3\farcm5, at a pixel size of 0\farcs05/pix.
The HRC images taken for NGC 1487 and the Antennae supernova
had a total field size of around 31''x31'', at a pixel size of
0\farcs027/pix.

For photometry, we use the photometric zeropoint determined by
\citet{DeMarchi04} and \citet{Sirianni05}.  The reduced $HST$ images
were combined with our Ks-band images in order to create the
two-colour-images shown in Figures \ref{antennaefig} and \ref{n14fig}.

\section{Analysis}\label{analysis}

\subsection{Velocity Dispersion Measurements}\label{estveldispsection}

\begin{figure}
\begin{center}
\psfig{figure=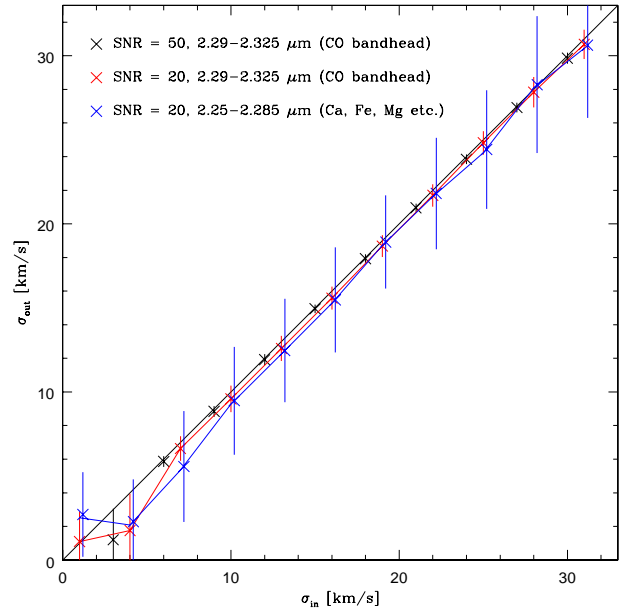,width=8.8cm}
\caption{Results of the fits of artificially broadened spectra,
which had noise added corresponding to signal-to-noise-ratios (SNRs) of 50 and 20, respectively
(where SNR$\approx$20 is the SNR obtained for most of our clusters). At our typical S/N, strong
deviations are only expected at less than half of the resolution of our
data (or $<$6-7 km s$^{-1}$).
\label{sigmafittests}}
\end{center}
\end{figure}

For each cluster spectrum, we estimated the width $\sigma$ of the 
broadening of the CO absorption 
features (assumed to be a Gaussian) which best fit the cluster
spectrum in the following way.  An appropriate stellar template
spectrum (described below) was broadened by Gaussian functions of
variable $\sigma$, ranging from 0 to a few 100 km/s, and shifted in
wavelength by radial velocities between 1400 and 1800 km/s.
We set the
starting point at  15 km/s and a radial velocity of 1600 km/s.  
The resulting set of
broadened templates were then compared with the cluster spectrum.  
The best fit was determined by evaluating $\chi^2$ and then searching for
the minimum of the function $\chi^2$(v$_r$,$\sigma$) using a simplex
downhill algorithm.
Radial velocities are given in Table \ref{velocitytable}, velocity
dispersions in Table \ref{massestable}.

\begin{figure*}
\psfig{figure=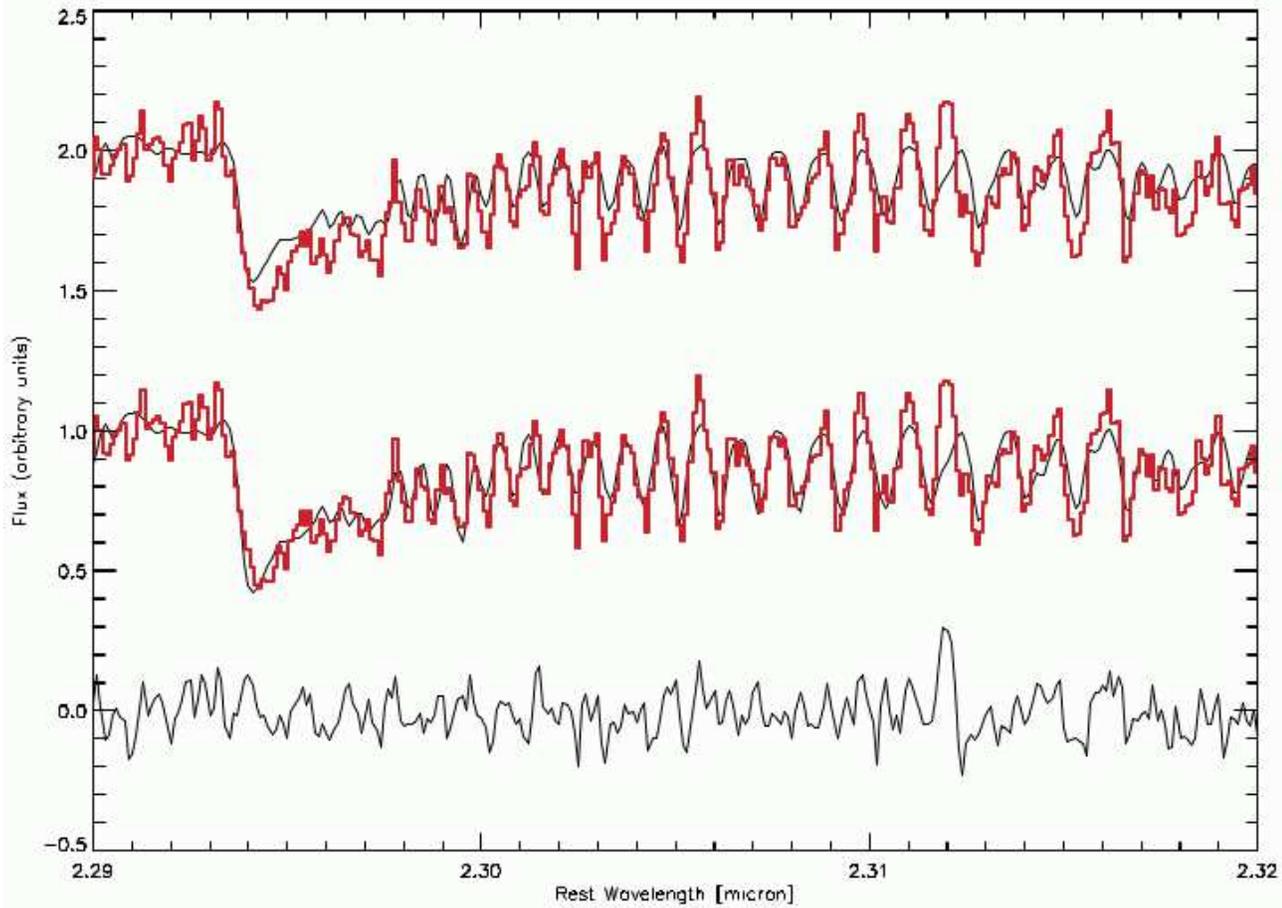,width=17cm,angle=270}
\caption{ISAAC spectrum of NGC 1487-1 with a good fit (using an M5I template
spectrum) with a velocity dispersion of $\sigma=13.7$ km/s as the bottom
spectrum, and a bad fit (11 km/s) at the top. The residuals are for
the good fit.\label{goodandbadfitISAAC}}
\end{figure*}

\begin{figure*}
\psfig{figure=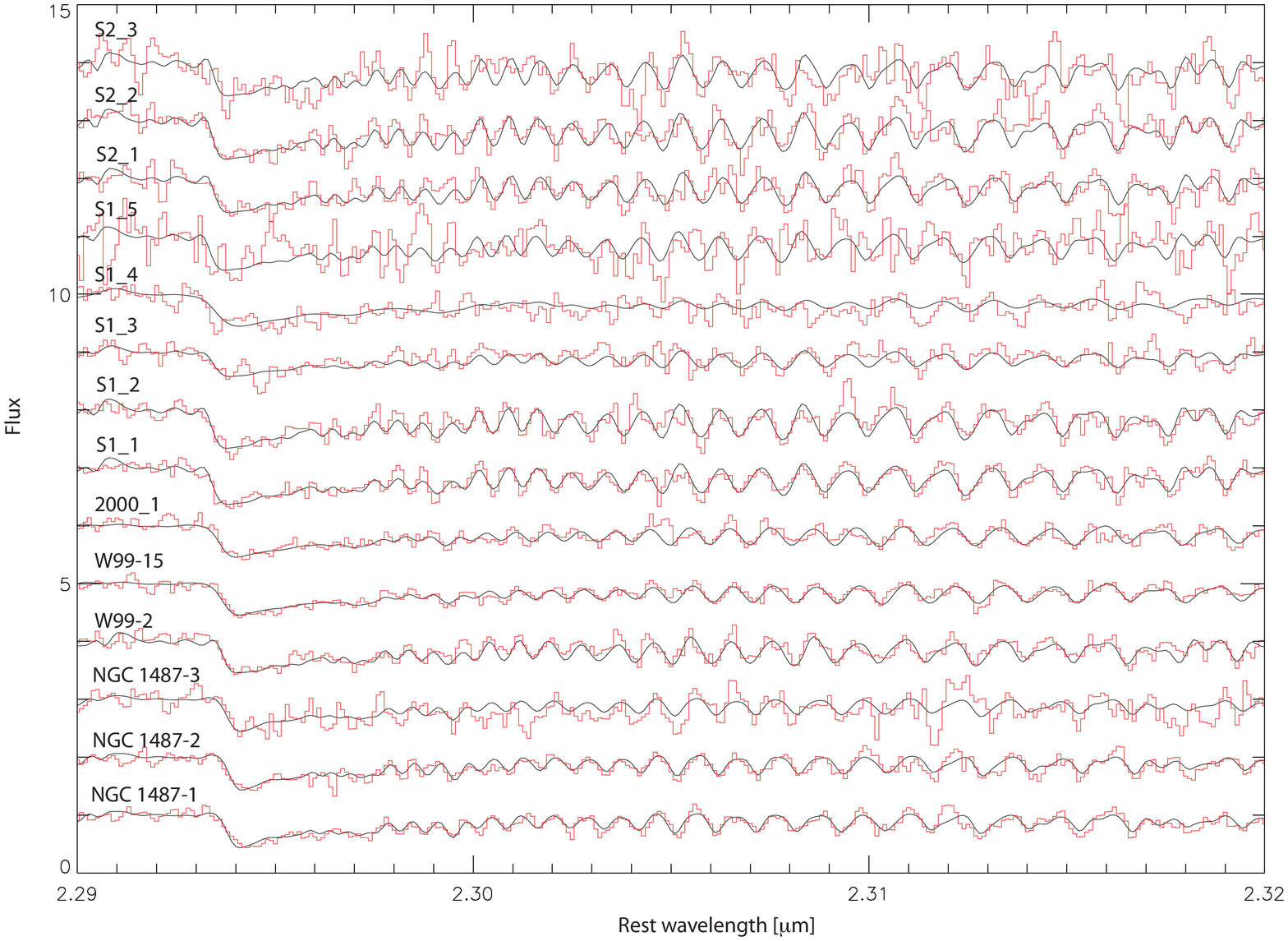,width=17cm}
\caption{All ISAAC spectra, overplotted with their best fits, with labels.
The region shown here generally provided the best fit.
\label{bestfitISAAC}}
\end{figure*}

It is important that the template spectrum be a good overall match in
terms of stellar features to the cluster spectrum.  For a star cluster
that formed $\sim$10 Myrs ago, late K through early M supergiant stars
are expected to provide the largest contribution to the 2.3$\mu$m
flux.  However population synthesis models \citep{Letal99} also show
that hot main sequence stars will make a non-negligible contribution
to the flux at this wavelength.  Since hot O and B-type stars have an
essentially featureless spectrum in the near-infrared, they only
represent a diluting continuum, which decreases the equivalent width
of the CO band-heads.  This has the effect of shifting the apparent
dominant stellar type towards higher effective temperatures.
Starting out with a template spectrum with weak CO features leads to
very low velocity dispersions, while the the opposite is true
if an M5I star (which has strong band-heads) is used, with substantial
differences in the results (a few km s$^{-1}$ to up to about 30 km
s$^{-1}$).  

We believe that no significant bias has been introduced through the
selection of a stellar template and/or the wavelength range which was
considered, because we have a large suite of templates which allowed
us to find a good match for each cluster. For a good match the best
fitting velocity dispersion was essentially independent of the
selected wavelength range, see \citet{Mengel02} for details.  While
we cannot rule out slight mismatches between template and cluster
spectra, particularly for the low-SNR clusters S2\_3 and S1\_5, these
should only have a slight effect on our velocity dispersion
measurements, and hence $M_{dyn}$ estimates.

The determination of the velocity dispersion from the optical echelle data
used the same procedure as for the ISAAC spectra, relying on the Calcium
Triplet around 8500\AA, but also using the MgI absorption feature at
8800\AA\ and other weaker metal absorption lines between 8400 and 9000\AA.

Clusters [W99]-2 and NGC 1487-1 were observed with both instruments,
ISAAC and UVES, in order to check the consistency of the final
velocity dispersion measurements. The agreement was excellent for the
Antennae cluster ($\sigma = 13.9 \pm 2.0$ km/s from ISAAC and $\sigma
= 14.3\pm 0.5$ km/s from UVES) and reasonable for NGC 1487-1 ($\sigma
= 12.0\pm2.0$ km/s and $\sigma = 15.4\pm2.0$ km/s), revealing no
strong systematic instrumental effect.

Our velocity dispersion measurements 
are summarized in Table \ref{massestable}.  We find a
range from about 7 km s$^{-1}$ to more
than 20 km s$^{-1}$.
For the optical echelle data the instrumental
resolution is only $\sigma_{instrument}$=3.2 km s$^{-1}$, ensuring that the cluster
line profiles are always well-resolved.  The situation for the near-IR
data is somewhat less straightforward, because the instrumental resolution is
$\sigma_{instrument}$=14.2 km s$^{-1}$, and some of our measurements
lie below this value. However, as shown in \citet{Mengel02} and Fig. \ref{sigmafittests}, results 
are reliable and reproducible down to approximately half this
value. 

Fig. \ref{sigmafittests} supports the assumption that the origin of this effect is
the large amount of signal contained in the CO bandheads.
We used our stellar spectra to create artificially broadened, noisy
spectra, and re-determined the velocity dispersion 100 times for each
sampled input velocity dispersion. We used two different wavelength
ranges of equal length, one including the CO bandhead, and one which
only includes only several metal absorption lines. The fit using the
CO region has a standard 
deviation which is $\approx$ five times smaller than that from the
other region. Mean fit values from the two different wavelength
regions start separating around $\sigma$=16 km/s, becoming quite
substantial below 10 km/s. The region around the CO bandhead
allows re-determination of the velocity dispersion down to around 6 km/s.
Therefore we believe that our measured velocity dispersions,
within their reported uncertainties, are valid, because they all
lie at or above the limit for a reliable CO fit.

\begin{table*}
\begin{center}
\caption{Radial velocities and J2000 coordinates of the Antennae clusters.\label{velocitytable}}
\begin{tabular}{lcccccc}
\hline
\hline
Cluster      & v$_r$ (observed)      & Heliocentric correction & v$_r$ (corrected) & RA              &       Dec \\
             & km/s                  & km/s                    & km/s              & hms             &                 \\\hline
$[$W99$]$2   & 1628.5 $\pm$0.5       & -9.8                    & 1618.7            & 12:01:50.46     &  -18:52:13.89   \\
$[$W99$]$15  & 1621.3 $\pm$0.3       & -9.0                    & 1612.3            & 12:01:55.43     &  -18:52:19.45   \\
S1\_1        & 1604.0 $\pm$0.2       & -8.1                    & 1595.9            & 12:01:55.20     &  -18:52:18.35   \\
S1\_2        & 1602.0 $\pm$0.2       & -8.1                    & 1593.9            & 12:01:55.26     &  -18:52:20.04   \\
S1\_3        & 1594.3 $\pm$0.3       & -8.1                    & 1586.2            & 12:01:55.49     &  -18:52:25.49   \\
S1\_4        & 1608.0 $\pm$1.2       & -8.1                    & 1599.9            & 12:01:55.05     &  -18:52:14.67   \\
2000\_1      & 1618.6 $\pm$0.2       & -9.0                    & 1609.6            & 12:01:55.29     &  -18:52:17.18   \\
S2\_1        & 1580.7 $\pm$0.2       & -8.6                    & 1572.1            & 12:01:55.54     &  -18:52:20.73   \\
S2\_2        & 1644.5 $\pm$0.4       & -8.6                    & 1635.9            & 12:01:51.90     &  -18:52:27.94   \\
S2\_3        & 1658.2 $\pm$0.2       & -8.6                    & 1649.6            & 12:01:53.13     &  -18:52:25.63   \\
S1\_5        & 1645.5 $\pm$2.0       & -8.1                    & 1637.4            & 12:01:54.31     &  -18:51:56.86   \\\hline
\end{tabular}
\end{center}
\end{table*}

Figures \ref{goodandbadfitISAAC} and \ref{bestfitISAAC} show some of the fits to the observed
cluster spectra. While Fig. \ref{goodandbadfitISAAC} shows a good fit in the middle (with the
corresponding residuals at the bottom), and a bad fit at the top, Fig. \ref{bestfitISAAC} shows the
best fits for all our spectra. In general, we obtained good and stable fits for the clusters,
even - surprisingly - for the low SNR spectra for S1\_5 and S2\_3.

\subsection{Cluster Sizes}\label{sizeestsection}

Sizes from ACS images were determined using the routine {\sl ishape},
implemented in the data reduction package {\sl baolab}, developed by
S. Larsen \citep{Larsen99}.  The clusters all appear slightly-to-well
resolved in the $HST$ images.  {\it ishape} convolves a user-provided
PSF with an analytic cluster profile, and determines the minimum
$\chi^2$ for a range of sizes using a simplex downhill algorithm.
{\it ishape} rejects pixels which deviate strongly from the median
value of pixels at the same radius outside a ``clean'' radius, which
we set to 3 pixels. The fit is performed out to a radius which we set
to 10 pixels for most clusters, 12 pixels for [W99]-2 and N1487-1
(which were the brightest clusters of each target, therefore the SNR
was sufficient out to a larger radius).  We have taken into account
the ellipticity which was assumed for the best fit model, and
determined the projected half-light radius which would correspond to a
circular model by using the average of r$_{min}$ and r$_{max}$. The
general validity of this approach remains to be verified, but is
currently justified by geometric considerations and some numerical
integrations \citep{LarsenManual}.  We converted the output FWHM from
{\it ishape} to a half-light radius by applying the appropriate
concentration-dependent conversion factor, as described in
\citet{Larsen01}. We convert these effective radii from arcseconds to
parsec by assuming distances of 19.3 Mpc (Antennae) and 9.3 Mpc (NGC
1487).

For the NGC 1487 ACS/HRC data, we created our PSF from archival data of a moderately
bright star (same filter and camera as the science data) which was obtained
for a completely different purpose (Proposal-ID 10198, PI: Wozniak).
The Antennae ACS/WFC data contained several foreground stars which we used
as PSF. Despite careful shifting-and-adding, a PSF created from more than two 
or three stars was always slightly broader than the original PSFs. Therefore
our PSF reference was created from only three stars.
For the Antennae ACS/HRC data, it was a lucky coincidence that the supernova
was located in the direct vicinity of cluster [W99]-2, thereby providing a
suitable PSF reference.

Our PSFs had the following characteristics: For the Antennae images, the
FWHM were 2.0 pix (0\farcs10) for F814W, 1.89 pix (0\farcs095) for F550M,
and 2.48 pix (0\farcs066) in the F555W filter, where the first two were using
the WFC, and the third the HRC. Only the F555W PSF showed a slight deviation
from a Gaussian profile, there was a slight (~5\% of the peak) increase in the count
level $\approx$3.5 pix north of the peak. FWHM of the PSFs used for the NGC 1487 images
were 2.58 pix (0\farcs07) in the F814W filter, and 1.97 pix (0\farcs053) in the F435W filter.
Apart from the strong Airy ring in the F814W PSF, they showed no peculiarities.

Best fits were typically obtained for King \citep{King62} profiles, with concentration
parameters between c = 5 and 300. For most clusters, we determined the best-fitting
half-light radius by running {\it ishape} for distinct values of c (5, 15, 30, 100, 300),
and additionally Moffat15, Moffat25 and Gaussian profiles. We did not generally do
a two-parameter fit (optimizing c and r$_{hp}$ at the same time), because we believe
that this increases the risk of getting trapped in a local minimum. But for two clusters
([W99]-2 and S2\_1) we did implement a two-parameter fit, and obtained satisfactory results
(the concentration for the best fit lay between the two best fixed-c fits).

The projected half-light radii which resulted from
the optimization in the two different filters are listed in Table \ref{sizestable}, together with
concentrations. The agreement between the filters was excellent for all the NGC 1487 clusters,
[W99]-15 and S1\_2, reasonable for [W99]-2 and S2\_2, and not very good 
for three clusters, S1\_5, S2\_1 and S2\_3, for reasons that are not clear. From visual inspection, the fit in
F550M looked much better for S2\_1, which is why we use this fit as the final value. But for
the two other clusters, all fits look quite reasonable, and we used averages (with rather
large uncertainties) as final values.
Cluster sizes range from $\approx1 - 8$pc, with a median size of 2.9~pc.
These values are fairly typical for young star clusters \citep[e.g.][]{Larsen04,Leeetal05}.

\begin{table*}
\begin{center}
\caption{Projected half-light radii of all clusters which do not have very obvious multiple components.
Sizes were measured in two different filters (where possible), in one additional filter for 
[W99]-2, and not at all for 2000\_1. The latter was assigned the same size and concentration as 
the two clusters in its vicinity (S1\_1 and S1\_2), because the three clusters appear comparable
in size in the K-band images.$^h$ was measured on an ACS/HRC-F555W image.\label{sizestable}}
\begin{tabular}{l|ccc|ccc|cc}
\hline
\hline
Cluster      &  r$_{hp}$(F814W) &   c(F814W)  & $\epsilon$ & r$_{hp}$(F550M) & c(F550M) & $\epsilon$ & r$_{hp}$    &   c    \\
             &  pc & r$_t$/r$_c$ & r$_{min}$/r$_{max}$  &    pc     & r$_t$/r$_c$ & r$_{min}$/r$_{max}$  & pc & r$_t$/r$_c$ \\\hline
$[$W99$]$2   &  9.4$\pm2$        & 30       & 0.92    & 7.3$\pm$2    & 300     & 0.87        &  8.0$\pm$1.5  & 150          \\
$[$W99$]$2$^h$&                  &          &         & 6.8$\pm$2    & 141     & 0.88        &               &              \\
$[$W99$]$15  & 1.4$\pm$0.2       & 300      & 0.91    & 1.5$\pm$0.4  & 300     & 0.85        &  1.4$\pm$0.2  & 300          \\
S1\_1        & 3.6$\pm$0.3       & 15-300   & 0.70    &              &         &             &  3.6$\pm$0.3  & 150          \\
S1\_2        & 3.6$\pm$0.5       & 300      & 0.69    & 3.6$\pm$0.2  & 15-300  & 0.74        &  3.6$\pm$0.4  & 300          \\
2000\_1      &                   &          &         &              &         &             &  3.6$\pm$1.0  & 300          \\
S2\_1        & 2.0$\pm$0.6       & 300      & 0.68    & 3.7$\pm$0.5  & 152     & 0.71        &  3.7$\pm$0.5  & 150          \\
S2\_2        & 2.0$\pm$0.5       & 300      & 0.75    & 2.9$\pm$0.4  & 100-300 & 0.66        &  2.5$\pm$0.5  & 300          \\
S2\_3        & 2.7$\pm$0.5       & 300      & 0.68    & 3.9$\pm$0.3  & 30-300  & 0.77        &  3.0$\pm$1.0  & 300          \\\hline
S1\_5        & 1.2$\pm$0.3       & 15-300   & 0.90    & 0.3$\pm$0.2  & 5-15    & 0.78        &  0.9$\pm$0.7  & 15           \\
\multicolumn{4}{c|}{} & r$_{hp}$(F435W) & c(F435W) & & \multicolumn{2}{c}{}\\\hline
NGC 1487-1   & 2.7$\pm$0.3       & 30       & 0.83    & 3.0$\pm$1.0  & 30      & 0.74        &  2.8$\pm$0.5  & 30           \\
NGC 1487-2   & 1.0$\pm$0.3       & 100-300  & 0.75    & 1.4$\pm$0.2  & 5-15    & 0.66        &  1.2$\pm$0.2  & 300          \\
NGC 1487-3   & 2.0$\pm$0.3       & 5-300    & 0.70    & 2.2$\pm$0.4  & 15-300  & 0.94        &  2.1$\pm$0.2  & 200          \\\hline
\end{tabular}								             	      
\end{center}									      								      
\end{table*}									      
										      
\begin{figure*}
\begin{center}
\psfig{figure=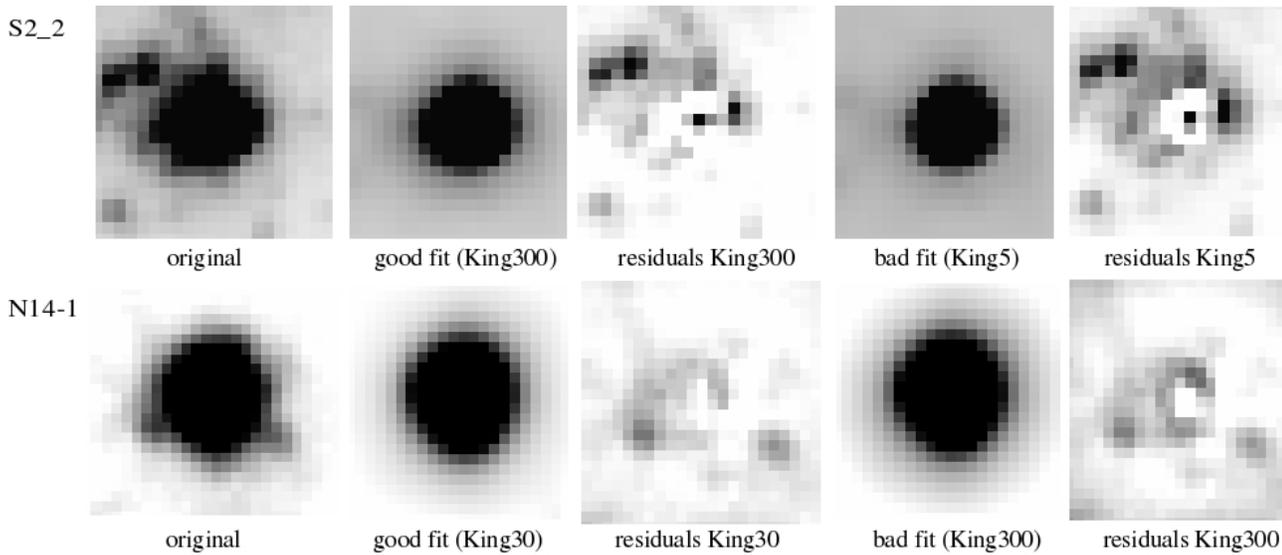,width=17cm,angle=270}
\caption{\label{clustersizefit} Original image, fit and residuals for two clusters
(S2\_2 and NGC1487-1) for the best and a bad fit each. Pixels deviating substantially
from the radial median are excluded from the fit and therefore expected to
show up in the residuals, for example the two fainter objects
in the top left corner of S2\_2. The better fits are characterised by a smoother
distribution at the central location of the cluster.
For each cluster, the images are shown with the same greyscale range.}
\end{center}
\end{figure*}

\begin{table*}
\begin{center}
\caption{Absolute extinction corrected magnitudes (distance
moduli of 31.41 and 30.13 for NGC 4038/4039 and NGC 1487, respectively), 
ages, velocity dispersion $\sigma$
(average value from ISAAC and UVES for [W99]-2 (14.0$\pm0.8$ and 14.3$\pm0.5$)
and NGC 1487-1 (16.3$\pm1.5$ and 15.4$\pm2.0$)).
Projected half-light radius r$_{hp}$.   
M$_{vir}$ is the virial mass determined from equation \ref{mvirequation} with the
1$\sigma$-uncertainties given in column $\Delta$M. M$_{ph}$ is the mass expected 
from Starburst99 models (version 5.0, 2005, \citep{Letal99, VazquezLeitherer05},
instantaneous burst with solar metallicity) 
for a cluster with the given absolute K-band magnitude and age.
  \label{massestable}.}

\begin{tabular}{lccccccccccccc}
\hline
\hline
Cluster      &  M$_K$(0)   & A$_V$  & Age         & $\sigma$     & r$_{hp}$     &   M$_{dyn}$  &  L$_K$/M & M$_{ph}$ & M$_{dyn}$/M$_{ph}$ &  $P/G$ \\
             &  mag        & mag    & [10$^6$yr]  & [km/s]       & [pc]         & [10$^6$M$_{\odot}$]  & [L$_{\odot}$/M$_{\odot}$]  & [10$^6$M$_{\odot}$] & & 10$^9M_{\odot}^2$/pc$^4$]  \\\hline
$[$W99$]$2   & -17.4$\pm$0.1   &  $\approx$0  & 6.6$\pm$0.3 & 14.1$\pm$1.0 & 8.0$\pm$1.5 & 3.0$^{+1.2}_{-1.0}$    & 64$^{+40}_{-23}$   & 2.7$^{+0.8}_{-0.5}$   & 1.0$^{+0.8}_{-0.4}$  & 1.9$^{+2.7}_{-1.2}$  \\
$[$W99$]$15  & -15.5$\pm$0.1   &  1           & 8.7$\pm$0.3 & 20.2$\pm$1.5 & 1.4$\pm$0.2 & 1.0$^{+0.3}_{-0.15}$   & 34$^{+10}_{-10}$   & 0.5$^{+0.1}_{-0.1}$   & 1.9$^{+1.2}_{-0.5}$ & 60$^{+50}_{-36}$  \\
S1\_1        & -15.7$\pm$0.1   &  4.6         & 8.0$\pm$0.3 & 12.5$\pm$3   & 3.6$\pm$0.3 & 1.0$^{+0.6}_{-0.4}$    & 38$^{+28}_{-16}$   & 0.7$^{+0.2}_{-0.1}$   & 1.5$^{+1.4}_{-0.8}$ & 2.9$^{+1.5}_{-1.4}$  \\
S1\_2        & -15.4$\pm$0.2   &  2           & 8.3$\pm$0.3 & 11.5$\pm$2.0 & 3.6$\pm$0.4 & 0.8$^{+0.4}_{-0.3}$    & 37$^{+33}_{-14}$   & 0.5$^{+0.3}_{-0.1}$   & 1.7$^{+1.4}_{-1.0}$ & 1.4$^{+1.1}_{-0.9}$  \\
S1\_5        & -14.8$\pm$0.1   &  2           & 8.5$\pm$0.3 & 12.0$\pm$3   & 0.9$\pm$0.6 & 0.4$^{+0.6}_{-0.3}$    & 46$^{+50}_{-21}$   & 0.3$^{+0.04}_{-0.05}$   & 1.4$^{+1.9}_{-1.0}$ & 103$^{+767}_{-94}$  \\
2000\_1      & -16.8$\pm$0.3   &  $\approx$10 & 8.5$\pm$0.3 & 20.0$\pm$3   & 3.6$\pm$1.0 & 2.4$^{+1.3}_{-0.9}$    & 46$^{+226}_{-30}$  & 1.7$^{+0.9}_{-0.6}$   & 1.5$^{+3.2}_{-1.2}$ & 16$^{+44}_{-14}$  \\
S2\_1        & -15.2$\pm$0.2   &  1.2         & 9.0$\pm$0.3 & 11.5$\pm$2.0 & 3.7$\pm$0.5 & 0.9$^{+0.5}_{-0.4}$    & 27$^{+29}_{-13}$   & 0.3$^{+0.12}_{-0.06}$   & 2.7$^{+2.5}_{-1.5}$ & 0.6$^{+0.6}_{-0.3}$  \\
S2\_2        & -15.3$\pm$0.1   &  0.5         & 9.0$\pm$0.3 &  9.5$\pm$2.0 & 2.5$\pm$0.5 & 0.4$^{+0.26}_{-0.17}$  & 72$^{+72}_{-33}$   & 0.4$^{+0.08}_{-0.03}$   & 1.0$^{+0.9}_{-0.6}$ & 3.5$^{+6.0}_{-2.1}$  \\
S2\_3        & -14.8$\pm$0.1   &  $\approx$0  & 9.0$\pm$0.3 &  7.0$\pm$2.0 & 3.0$\pm$1.0 & 0.25$^{+0.32}_{-0.17}$ & 70$^{+168}_{-42}$   & 0.24$^{+0.07}_{-0.03}$   & 1.0$^{+1.7}_{-0.8}$ & 0.7$^{+3.3}_{-0.5}$  \\
NGC 1487-1   & -14.2$\pm$0.1   &  0.6         & 8.4$\pm$0.5 & 13.7$\pm$2.0 & 2.3$\pm$0.5 & 1.2$^{+0.7}_{-0.5}$    & 8.1$^{+6.5}_{-3.3}$ & 0.15$^{+0.05}_{-0.02}$  & 8.2$^{+6.3}_{-4.5}$ & 0.80$^{+1.7}_{-0.5}$  \\
NGC 1487-2   & -14.2$\pm$0.1   &  1.0         & 8.5$\pm$0.5 & 11.1$\pm$1.8 & 1.0$\pm$0.3 & 0.2$^{+0.11}_{-0.08}$  & 48$^{+36}_{-19}$   & 0.16$^{+0.05}_{-0.02}$  & 1.3$^{+1.0}_{-0.7}$ & 26$^{+46}_{-16}$  \\
NGC 1487-3   & -13.4$\pm$0.3   &  0.5         & 8.5$\pm$0.5 & 14.3$\pm$1.0 & 1.8$\pm$0.3 & 0.6$^{+0.20}_{-0.13}$  & 7.7$^{+5.2}_{-3.3}$  & 0.076$^{+0.04}_{-0.03}$ & 8.2$^{+9.6}_{-4}$ & 0.6$^{+0.3}_{-0.4}$  \\\hline
\end{tabular}
\end{center}
\end{table*}

\subsection{Clusters with multiple components}\label{multiples}

For essentially all of our clusters, faint additional point sources (or slightly 
resolved objects) can be detected within the slit width of 0\farcs3 
(see Fig. \ref{Clusterblowups}). While in most cases, these 
objects only contribute a few percent of the flux of the primary, we
have indentified several objects where the impact of the nearby cluster
is possible but difficult to quantify, or is obvious and perhaps severe. Two of the clusters
were discarded from further analysis for this reason: S1\_3 and S1\_4
which had been only secondary targets in slit position S1. The multiple
components which show up in the ACS image (see Fig. \ref{Clusterblowups}) 
are also obvious, even though not as well resolved, in the
ISAAC K-band image as extended and elongated PSFs.

\begin{figure*}
\begin{center}
\psfig{figure=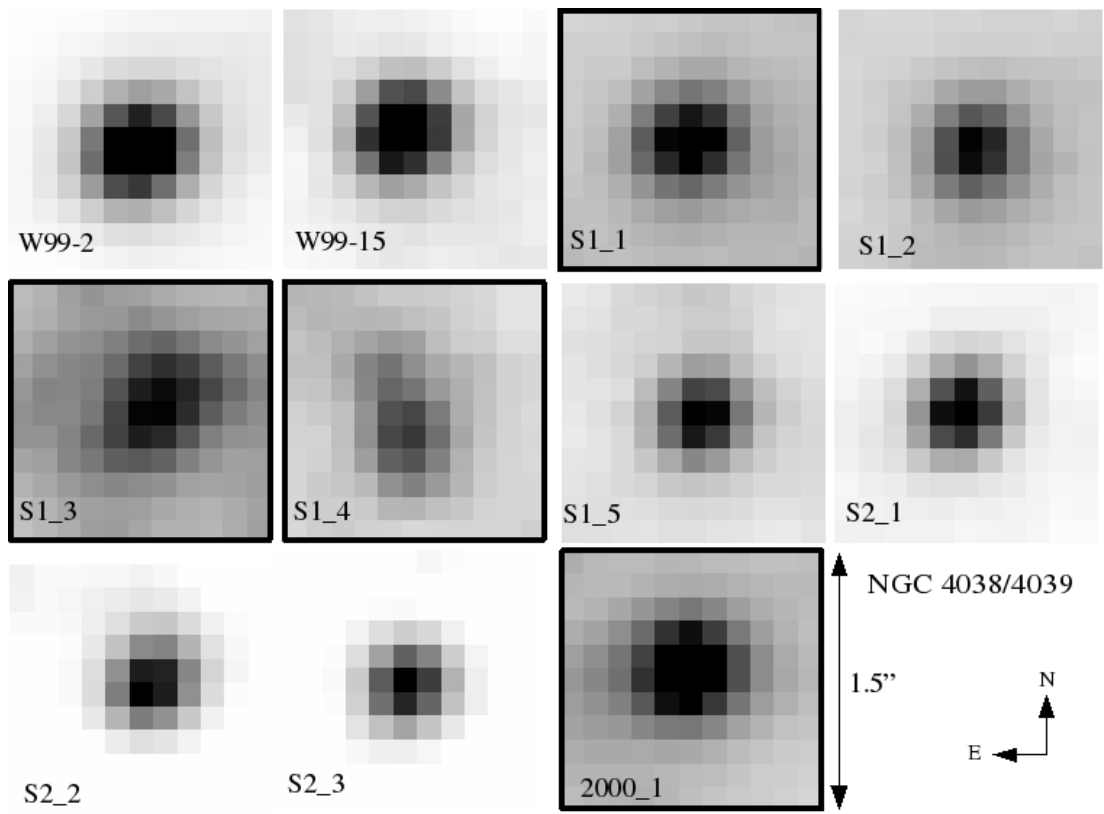,width=17cm,angle=270}
\caption{VLT-ISAAC Ks images of Antennae clusters included in our spectroscopy
slits. The clusters
which are marked with boxes are those which have an obvious strong neighbour
or which consist of multiple objects. This could have an impact on the
measured velocity dispersion, and the two most obviously affected
clusters (S1\_3 and S1\_4) were discarded from further analysis for this
reason.\label{Kclusterblowups}}
\end{center}
\end{figure*}

This is not the case for the other two Antennae clusters with multiple components
in the F814W image: For both, S1\_1 and 2000\_1, the PSF looks symmetric, and
there is no obvious indication of multiplicity. This could mean that
the companion clusters are bluer than the main component and therefore
their contribution in K-band, where we estimate both the velocity dispersion
and the photometry of the cluster, may be negligible. 
Nevertheless, we have marked these clusters in all our analysis plots,
even though their properties do not turn out to be unusual in any obvious way.

N1487-3 is a candidate which is expected to have
suffered some impact on its measured velocity dispersion:
While we think that photometry took reasonable account of the multiplicity,
and only uses the flux from the primary component to estimate the magnitude
and photometric mass of the cluster, some flux
from the companion cluster leaked into the slit during spectroscopy and cannot be removed.
This is the case at a low level during the major fraction of the integration time
where the seeing was excellent (FWHM 0\farcs3-0\farcs4 in K-band), and more 
pronounced during the $\approx$30\% of the integration time where the seeing
was larger (FWHM 0\farcs5-0\farcs6 in K-band). The distance between the two 
components is 0\farcs5.

We believe that
we see the impact of an erroneously high velocity dispersion
(caused by the different radial velocity of the companion
cluster) in the very high ratio of M$_{dyn}$/M$_{ph}$.

\begin{figure}
\begin{center}
\psfig{figure=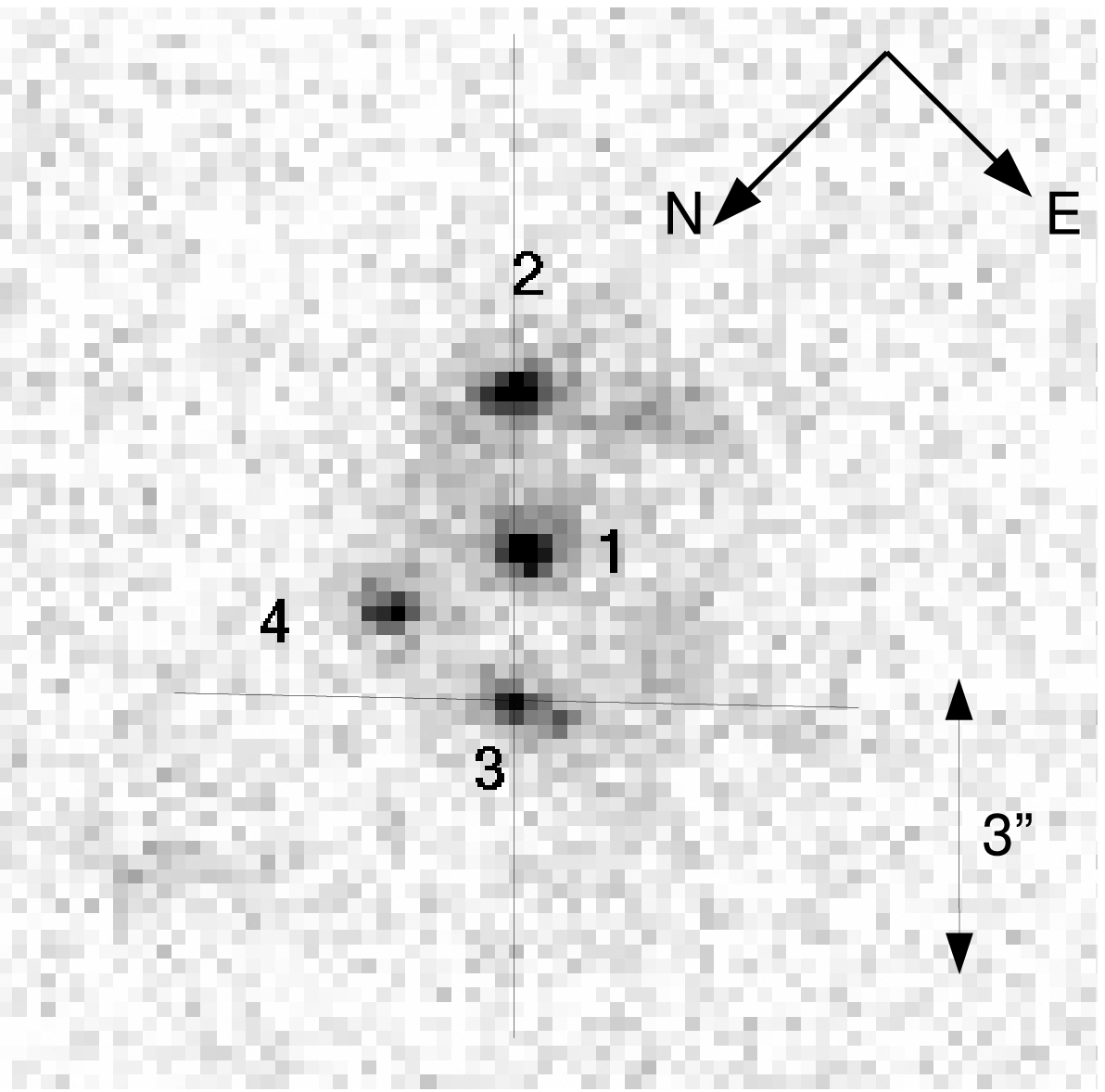,width=8.8cm}
\caption{VLT-ISAAC Ks acquisition image of NGC 1487.
It shows that the two components of Cluster 3 could be
well resolved. The slit was positioned on the brighter
component. Nevertheless, some flux from the fainter component
entered the slit.\label{N1487_acq}}
\end{center}
\end{figure}

\subsection{Dynamical Cluster Masses}\label{dynmassesubsection}

The dynamical mass of each cluster is estimated using 

\begin{figure}
\psfig{figure=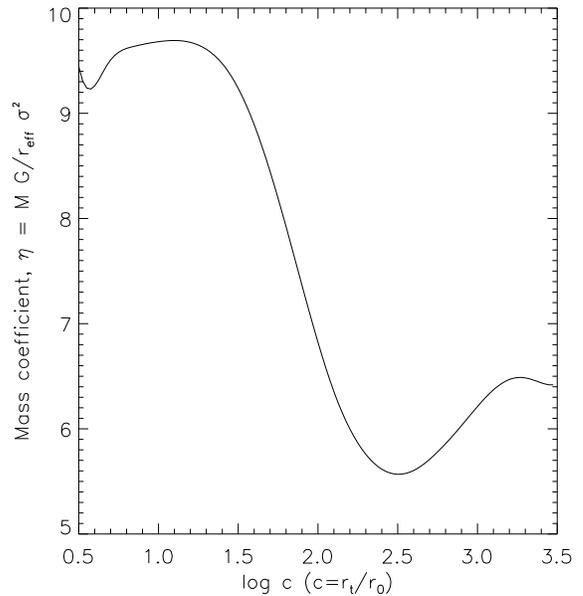,width=8.8cm}
\caption{Variation of the mass coefficient, $\eta$, as a function of 
the cluster concentration, c ($c = r_t / r_0$). $r_0$ is the King
radius, and $r_t$ is the tidal radius of the King model.
Our clusters cover the range from log(c) = 0.7 - 2.5. \label{etavsc}}
\end{figure}

\vspace{-0.2cm}
\begin{equation}\label{mvirequation}
M_{dyn}=\frac{\eta F \sigma^2 r_{hp}}{G}
\end{equation}
\vspace{-0.2cm}

where $\eta = \eta(c)$ is a factor that depends on the distribution of the
stellar density with radius (described below),
$\sigma$ is the stellar velocity dispersion, r$_{hp}$ is the projected half-light
radius, and G is the gravitational constant.  The function $\eta$ depends on both
the cluster concentration, $c$, and on the mass-to-light ratio as a function of
radius.  
We assume that it is justified to split these two dependencies into two separate
parameters: $\eta = \eta(c)$ and F = F(t).
Figure \ref{etavsc} shows the variation of  $\eta$ with concentration c.
To generate the relationship between the coefficient necessary to estimate the mass
of the clusters, $\eta$, and the ratio of the tidal radius to the core radius
(the concentration parameter, c) for a King model (Fig. \ref{etavsc}), we solved 
numerically Poisson's equation yielding the density distribution for a variety
of W$_0$ \citep[which is the central potential divided by the velocity dispersion
and characterises a King model; see e.g., equation 4-131 from][]{binneyandtremaine87}.
Integrating the density profile with radius provided the mass coefficient, $\eta$.

For all clusters in NGC 4038/39 and NGC 1487, $\eta$ varied between 5.6 and 9.7.

The factor F = F(t) describes how $\eta$ varies if the mass-to-light ratio
varies as a function of radius. 

As described in \citet{Fleck06}
their models indicate that the first 10 Myrs of cluster evolution - at
least for dense clusters like those in our sample - result in a steep increase in the
factor $\eta$ in Eq. \ref{mvirequation}, and a more gentle increase
after that.  This effect is caused by mass segregation, which leads to
a decrease in half-light radius, while the total mass and the
half-mass radius are largely unchanged - i.e. the mass-to-light ratio
varies with radius.  Even though it is expected to depend on several
parameters how strongly a cluster segregates (density, IMF, upper mass
cutoff, initial radius, number of stars), and only the density can
be determined a priori, we think that it should be expected
for our very dense, 10 Myr clusters.  

We applied an average
factor of F = $\eta_t / \eta_0$ = 1.3  (derived from Fig. 14 in
\citet{Fleck06}) to all our clusters.
While mass segregation is expected theoretically, it should be
noted that our barely resolved clusters make it impossible
to determine observationally
if they have undergone mass segregation or not, because statistics and
crowding will cause the highest-mass stars to appear mass segregated
in any strongly centrally concentrated cluster \citep{Ascenso2008}.   

The dynamical models do not take into account the contribution of
stellar binary orbital motion to the velocity dispersion; 
however this contribution is expected
to be negligible, due to the large masses of the clusters studied 
here \citep{Kouwenhoven07}.

\subsection{Photometric Ages and  Masses}\label{photmassesubsection}

In order to estimate the mass of each cluster based on our imaging
data, we use our K-band images, since these suffer from significantly less
extinction than the $V$ band.
We performed aperture photometry in two different ways:

For NGC 4038/4039, we used a curve-of-growth technique to 
determine the total magnitudes for a large number of clusters, 
resorting to magnitudes from fixed (small)
apertures plus an aperture correction only in cases where the
nearest neighbouring cluster was closer than a certain limit \citep{Mengel05}.

\begin{figure}
\begin{center}
\psfig{figure=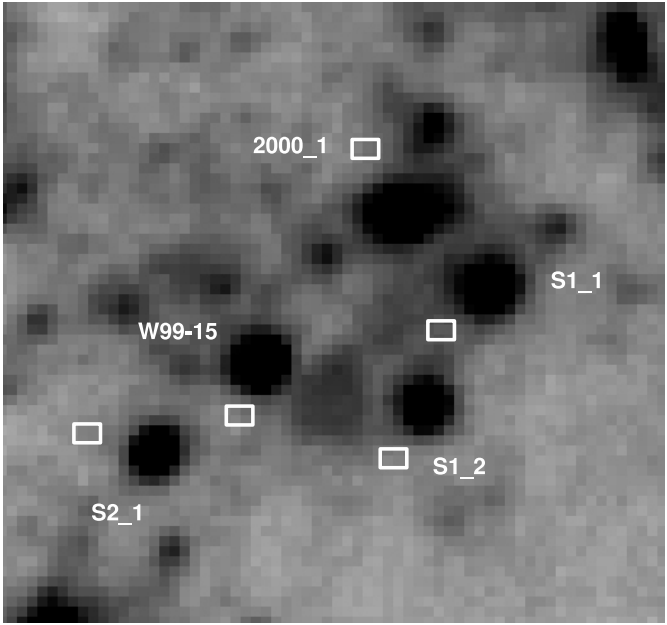,width=8.8cm}
\caption{Blowup of the densely populated overlap region in
the Antennae. In order to enhance the background, we chose a
logarithmic scaling. The rectangles indicate regions which we
selected manually as a reasonable assumption for the background
level at the location of the nearby cluster.
\label{Antclusterbackground}}
\end{center}
\end{figure}

Since some of our clusters are located in regions of backgrounds
which show a gradient (see Fig. \ref{Antclusterbackground}), 
and only a small number of clusters needs
to be treated, we additionally used a more manual approach
(which was the only technique used for the NGC 1487 clusters):
We chose an aperture size individually for each cluster where the
signal from the cluster was low enough to be comparable to the
noise, and selected a background region manually to be representative
of the background expected at the cluster location. Aperture sizes here
ranged from 1\farcs2 to 1\farcs8, roughly 3 to 4.4 times the FWHM of
the PSF in the Antennae images. For NGC 1487, an aperture size of
3\arcsec\ was also roughly four times as big as the FWHM of the images.

For the Antennae clusters, the two techniques gave identical
results, except for three clusters: S1\_1 is fainter by 0.1 mag using 
the manual technique, S1\_2 is 0.2 mag brighter. Differences for those
two clusters could be expected, because they are located right on the
edge of a variation in background intensity. S2\_1 is also brighter
by 0.2 mag using the manual approach, and here the reason is not
obvious. In all three cases, we used the manually determined value,
but increased the uncertainty estimate in the photometry to account for
the differences from the two methods.

Since the diffuse light in NGC 1487 is smooth, and none of the clusters
have confusing neighbouring clusters, we estimate that the photometry for
clusters NGC 1487-1 and NGC 1487-2 is reliable.

For cluster NGC 1487-3, the K-band photometry is
complicated by the fact that there is a second, fainter
cluster so close by that the two are only marginally separated in our
near-IR images. The large
uncertainty in the photometry is due to confusion from this neighboring source. 
To assign a K-band magnitude to the brighter cluster, we used an aperture that includes both
clusters and assumed that the relative brightnesses of
the two clusters is the same in I-band and K-band. The 
ratio of the peak counts in the acquisition image
shown in Fig. \ref{N1487_acq} supports this assumption. 

The absolute K band magnitudes range from -13.4 mag to -17.4
mag. The absolute extinction corrected K-band magnitudes are
converted to M$_{ph}$ by comparison with the absolute K-band 
magnitude predicted by a Starburst99 model for a 10$^6$ M$_\odot$
cluster (instantaneous burst, Kroupa IMF, solar metallicity) of
the same age. Resulting photometric masses are listed in Table~\ref{massestable}.

All clusters are fairly massive, at least
compared to young star clusters in the Milky Way. For interacting galaxies, these
luminosities, corresponding to photometric masses between
8$\times10^4$M$_\odot$ and 4.5$\times10^6$M$_\odot$, are not
unusual. Furthermore, we selected some of the most massive clusters
specifically, because, at comparable ages, they are the most luminous
and therefore most easily accessible to high resolution spectroscopy.  

The fact that the clusters in NGC 1487 are generally fainter and less
massive than those in the Antennae is expected from simple statistical
considerations. NGC 1487 has a relatively low number of clusters compared
to the Antennae -- only 1-10\%.  Moreover, because we targeted clusters
that are amongst the brightest in each galaxy, the likelihood that a
system like NGC 1487, produces one or more clusters which are comparable
to the brightest clusters in the Antennae, is comparatively low
\citep[e.g.][]{Whitmore2000, Whitmore07}.

\begin{figure*}
\begin{center}
\psfig{figure=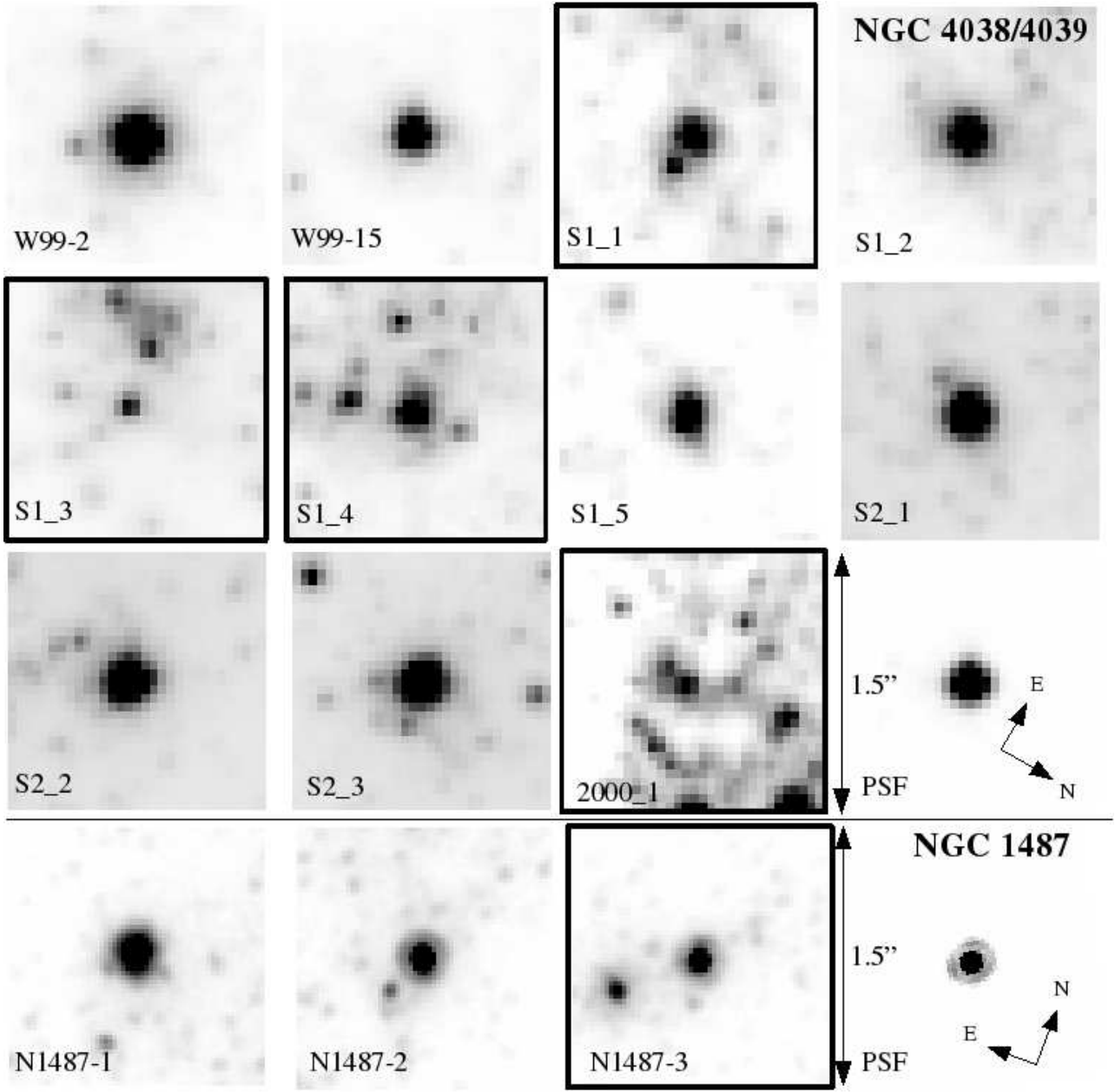,width=17cm,angle=270}
\caption{HST-ACS/WFC F814W images of the clusters used for our analysis. The clusters
which are marked with boxes are those which have an obvious strong companion
or which consist of multiple objects. This could have an impact on the
measured velocity dispersion.
\label{Clusterblowups}}
\end{center}
\end{figure*}

Extinction turned out to be relevant, at least in K-band, only for two
clusters, S1\_1 (A$_V$=4.6mag) and 2000\_1 (A$_V \approx$ 10 mag). For all the
other clusters, A$_V$ is below 2~mag (with the uncertainties 
$\Delta$A$_V \leq$ 0.5 mag), which translates to A$_K$ below
$\approx$ 0.2mag, making potential uncertainties of 
$\Delta$A$_K < 0.05$ mag small in relation to the overall
photometric uncertainty.

We estimate the age of each cluster by comparing the broadband filter
measurements, the CO(3-1) bandhead, and Br$\gamma$ and Calcium Triplet
(CaT, from UVES spectroscopy) equivalent widths  with the predictions of
the Starburst99 \citep{Letal99,VazquezLeitherer05} evolutionary
synthesis models.  

For all clusters, we determined a best fit age between 8 and 9 Myrs. For clusters
with ages $\approx$8.5 Myr, an alternative solution exists at 10.5
Myr for all cluster properties which are influenced by red supergiants
(M$_{K}$,W$_{CO}$,W$_{CaT}$, V-K, etc.) because the quantities are double
valued at around this age. Since M$_K$ is the same for both ages, all our
conclusions remain unchanged, because M$_{ph}$, L$_K$/M$_{dyn}$
etc. are unaffected by this uncertainty in age.

\section{Discussion and Implications}\label{discussion}

%\subsection{Cluster dissolution}\label{dynamics}

An underlying assumption of any mass estimate based on the velocity
dispersion is that the cluster is self-gravitating (i.e., bound).
The age of our clusters, $\approx10$~Myr, is much
older than their estimated crossing times ($t_{cross} \sim r_{hp}/\sigma$)
of $1-$few $\times$10$^5$ yrs, indicating that they have
already survived for 20-50
crossing times.
However, recent results suggest that a large fraction of clusters
becomes unbound and disperses within $\approx10-20$~Myr due to the removal
of interstellar material; therefore it remains possible that the
clusters studied here may not be in virial equilibrium.  If cluster
stars are dispersing, then this extra-virial motion will lead to a 
measured velocity dispersion  which is higher than would be measured
for a bound cluster of similar mass.

\begin{figure}
\begin{center}
\psfig{figure=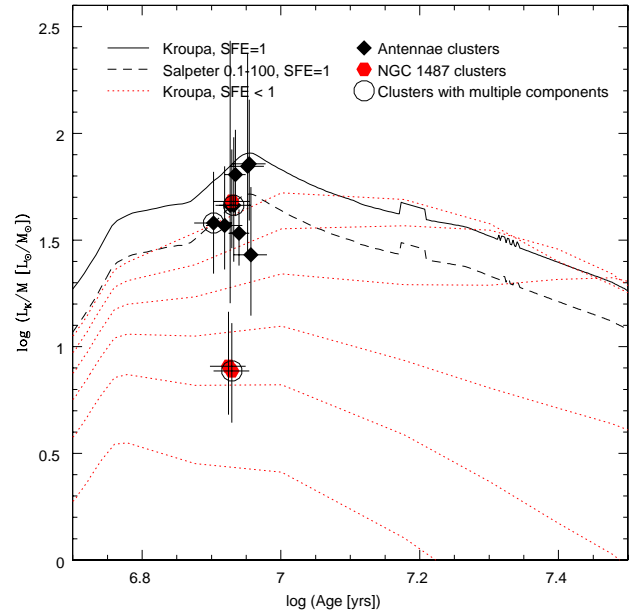,width=8.8cm}
\caption{This is a modified version of a
figure from \citet{GoodwinBastian06}; their Figure
5. It shows L/M for a Starburst99 (v.5.0, 2005) SSP with a Kroupa IMF, 
instantaneous burst, solar metallicity (solid, black). Dotted (red)
take into account the  Goodwin \& Bastian model results
(Bastian, priv. comm.) for clusters with 
effective star formation
efficiencies (eSFEs) below 100\% (lowest line: 10\%, top: 60\%). 
The locations of our clusters are indicated (clusters with
identical ages have been offset by 0.1 Myr in order to be able
to see which error bar belongs to which cluster).
Those which end up between the lines according to eSFEs of 20\%
and 50\% are discussed as dissolution candidates in the text,
and correspond to ratios of M$_{dyn}$/M$_{ph} \geq 2.5$.
\label{Clustersurvival}}
\end{center}
\end{figure}

One way to assess whether our clusters are gravitationally bound or
show evidence for non-virial motion is to compare the $L_K/M$
determined from velocity dispersion measurements with those predicted
by population synthesis models.  In essence, this is a comparison of
the dynamical and photometric masses.  A cluster which has $L_K/M$
(based on dynamical measurements) which is lower than the photometric
estimates (from the age of the
cluster and its measured stellar light), can point to non-virial motion
resulting from an expanding, dissolving cluster.  In
Figure~\ref{Clustersurvival} we compare our estimated cluster ages and
L$_K$/M for the Antennae and NGC~1487 clusters, with the predictions
for an instantaneous burst, solar metallicity model from Starburst99
(Leitherer et al. 1999).  We show predictions for L$_K$/M assuming two
different IMFs: a Kroupa IMF (solid line) and a Salpeter IMF with
$0.1~M_{\odot}$ and 100~$M_{\odot}$ lower and upper mass cutoffs
respectively (dashed line).  Within the measurement uncertainties, the
measured properties agree with the model predictions for all but two
clusters in NGC~1487 and one cluster in NGC~4038/4039.  
This would still be the case if we had assumed the distance to the
Antennae which was determined from the tip of the red giant branch \citep{Savianeetal08}
to be substantially lower than our assumed value, 13.3 rather than 19.3 Mpc.
Since this lower distance would affect both estimates (lowering M$_{ph}$ by
a factor 2, and lowering the cluster sizes and hence M$_{dyn}$ by a factor
1.45), the net effect would be that the ratio of M$_{dyn}$/M$_{ph}$ needs
to be corrected by a factor 1.38 (correspondingly decreasing L$_K$/M by a
factor 0.73). In general, this would still lead to a good 
correspondence between the photometric and dynamical estimates, and would
leave the conclusion the same:

The good agreement between evolutionary synthesis models applied to
our clusters, in comparison with dynamical masses, suggests that
there is no strong variation in the IMF for all but three clusters in
our sample, and that they have likely survived the gas removal phase
as bound stellar systems.

The two NGC~1487 clusters (red stars) and one Antennae cluster 
(S2\_1) which are offset below the model
predictions have dynamical mass estimates which are significantly higher
than the photometric ones.  One possible explanation is that these clusters
have an IMF which is significantly steeper than Kroupa/Salpeter.  There is little
direct evidence for such an interpretation, and we believe that it is much
more likely that these are clusters caught in the act of dissolving.
The efficiency with which a cluster forms stars will impact the
probability that it survives the expulsion of its natal gas.  
For example, clusters which form stars at lower efficiencies end up with
fewer bound stars (i.e. shallower potential wells) relative to the
left-over gas from formation.  Such clusters, as momentum input from the massive 
stars expels the gas, have a lower probability of remaining
bound than a cluster that formed more stars and had less remaining
gas.  \citet{GoodwinBastian06} explored the connection between
star formation efficiency and cluster dissolution by simulating
the N-body dynamics of a cluster after the expulsion of gas.
Using their results (Bastian, priv. comm.), we plot the
$L_K/M$ ratios for a Kroupa IMF, solar metallicity (Starburst99 models) as the 
red dotted lines in Figure~\ref{Clustersurvival}
for the following effective star formation efficiences (eSFEs, defined
as a measure of how far the cluster is out of virial equilibrium after
gas expulsion): 
60\%, 50\%, 40\%, 30\%, 20\%,
and 10\%, starting from the top. eSFEs $\approx40$\% and higher
are predicted to result in stable clusters
after $20-30$~Myr, even though many clusters, at least the
three between the 40\% and the 60\% lines, may lose a substantial
amount of mass (\cite{GoodwinBastian06}).  

\citet{BaumgardtKroupa2007} ran a grid of models with larger parameter
space, varying star formation efficiency (SFE, defined in the normal
way as the ratio of stellar mass over mass of stars and gas), gas 
expulsion time and tidal field. Even though a direct
comparison to the \citet{GoodwinBastian06} results is difficult,
some general conclusions are the same in both models: All clusters,
even the ``survivors'', expand initially, and almost indistinguishably.
After 10-20 Myr, the dissolving clusters continue to expand, while those
with a sufficiently high SFE re-contract. A more gradual gas expulsion
than the instantaneous expulsion assumed by \citet{GoodwinBastian06} makes
it easier to remain bound.
 
Two clusters in in NGC~1487 and one in the Antennae, in  Fig. \ref{Clustersurvival},
lie in or very close to regions where cluster dissolution is expected
from the \citet{GoodwinBastian06} models.
We consider two of them candidate dissolving
clusters (which is particularly interesting because of their high
mass - even though infant mortality is claimed to be mass
independent, high-mass clusters are usually intuitively considered
more stable against dissolution). Variations in the initial conditions,
for example longer gas expulsion times or tidal fields, 
as explored by \citet{BaumgardtKroupa2007}, would shift the boundary
between dissolving and surviving clusters in this plot down or up,
respectively.

The third cluster is NGC~1487-3, which is likely to have suffered
effects of cluster multiplicity (see Sect. \ref{multiples}).

One of the main results of this work is that most of the clusters in
our sample appear to have survived, as bound stellar systems, the gas
removal phase which occurs during the life of every cluster.  It is
important to note that, in light of many recent works which show that
many or most clusters (roughly 50\% to 90\%) probably do not survive
the earliest phases of evolution, our study targets clusters which are
likely to have survived this phase.  After $\approx10$~Myr other
mechanisms will continue to unbind clusters.  If essentially all
clusters which reach an age of $\approx10$~Myr in the Antennae are
marginially bound or bound at this point, this would imply a very large number of young
globular clusters.  
However, \citet{Fall05} show that, at least statistically,
star clusters continue to get disrupted approximately
independent of mass, out to an age of $\approx$100 Myr.

Evaporation of stars resulting from two-body
relaxation will eventually disrupt a number of lower mass clusters
over a Hubble time. Clusters with current masses $\geq
\mbox{few}\times10^5~M_{\odot}$ would likely survive this process over
a Hubble time, assuming the typical evaporation
rate of $\mu_{ev}$ = 1 - 2 $\times$ 10$^{-5}$ M$_{\sun}$ yr$^{-1}$.
Such a rate is plausible as it reproduces the observed turnover in the
mass function of globular star clusters in many galaxies
 \citep[e.g.,][]{FallZhang01, Watersetal06, Jordanetal06}.  

Spectroscopic studies like the one presented here require enormous
amounts of telescope time, and still result only in very few spectra.
It would be much more efficient if a cluster population
could be separated into ``survivors'' and ``dissolvers'' from
(high resolution) imaging alone, because this would give a better handle 
on the infant mortality rate.
Obviously, constraining the infant mortality rate is of immense relevance 
for the whole issue of star formation, because the
currently cited cluster destruction rates range in impact from
``cluster formation is an interesting, but not very important mode
of star formation'' (for destruction rates of a few tens of percent)
to ``essentially all stars formed in clusters'' (for destruction rates
of $\approx$90\% per decade).

\begin{figure}
\begin{center}
\psfig{figure=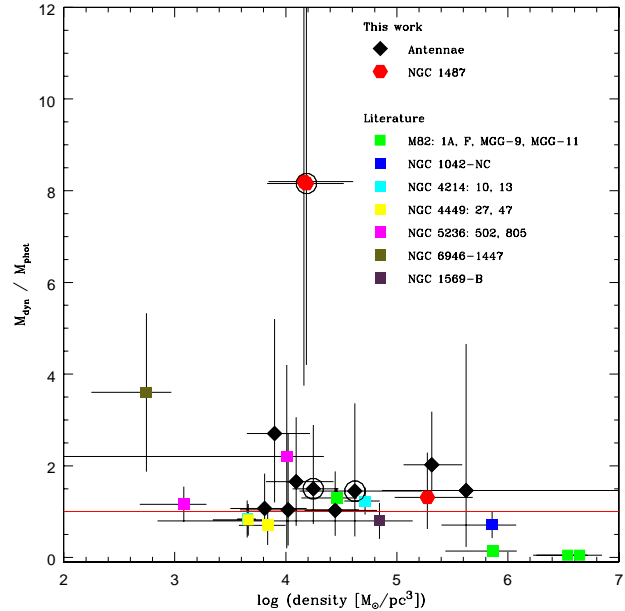,width=8.8cm}
\caption{The ratio of dynamical mass over photometric mass vs. half mass density $\rho_h$ (taken as the log on the X-axis). 
High values ($\geq 2.5)$ of M$_{dyn}$ / M$_{ph}$, which may indicate dissolving clusters, seem more common at low densities. Symbols
are described in the legend, diamonds and hexagons represent our data (with the open
circles marking clusters which may have been influenced by nearby clusters), squares 
are taken from the literature:
M82-F: \citet{McCrady05, McCrady07}, M82-A1 (1a, respectively): \citet{Smith06, McCrady07},
M82 MGG-9 and MGG-11: \citet{MGG03}, NGC 1042-NC: \citet{Boeker04a, Boeker04b, deGrijsetal05},
NGC 4212-10, 4214-13, NGC 4449-27 and NGC 4449-47: \citet{Larsenetal04}, NGC 5236-502, 5236-805: \citet{LarsenRichtler04},
NGC 6946-1447: \citet{Larsenetal04, Larsenetal06}, NGC 1569-B: \citet{Larsenetal08}.
\label{massratiovsdensity}}
\end{center}
\end{figure}

One expectation for expanding, unbound clusters is that as they expand, 
their internal density should decrease. Note, however that it is impossible to determine from
the size and/or concentration of these young clusters alone if
they are bound or dissolving: Even though unbound stars, leaving
the cluster with escape velocities of tens of km/s reach distances
of tens or hundreds of pc in a few Myr, the half-light radius of
the cluster is not immediately affected severely - even after 20 initial
crossing times, the half-mass radius of a dissolving cluster is only
40\% larger than that of a surviving cluster \citep{BaumgardtKroupa2007}. 
An alternative explanation for clusters having low density
is simply that they formed in a low-density environment,
which is expected to lead to smaller SFEs. In any case, we might expect 
our clusters with non-virial
motions to have low stellar densities as well.  In
Fig.~\ref{massratiovsdensity}, we show the estimated half-mass
density $\rho_h$ (with $r_{hp}$ and $M_{ph}$ as input) for the clusters 
versus the ratio dynamical to
photometric mass.  This figure shows that our two dissolving
cluster candidates (i.e. those with high ratios of
$M_{dyn}/M_{phot}$) also appear to have low stellar densities.  

We have included in Fig.~\ref{massratiovsdensity} all the data from the
literature where the three parameters M$_{dyn}$, M$_{ph}$ (assuming a
Kroupa IMF) and r$_{hp}$ were provided or could easily be deduced.
Most of the clusters in other publications are considerably older than
10 Myr, therefore it is not surprising that they do not show indication
of cluster expansion. But the two literature clusters where the 
dynamical mass exceeds the photometric mass by more than a factor two
(NGC 6946-1447 and NGC 5236-805) confirm the trend shown in our clusters,
since they also have low densities.

Clusters with high M$_{dyn}$/M$_{ph}$ likely have low values of
$\rho_h$ because they are expanding and thus the high values are  due to
dynamical evolution. Therefore clusters with ages around 10 Myr which have
low densities are excellent candidates for clusters in the process
of dissolving, although clearly some fraction of low density clusters
appear to be bound at this age as well.  Low density clusters which are gravitationally
bound, such as found in the Milky Way (outer globular clusters, \citet{Harris96}), the
Magellanic Clouds \citep{vandenBergh91}, and in nearby spirals and lenticular galaxies
\citep{ChandarWhitmoreLee04, LarsenBrodie00, Peng06}, may survive longer than their higher density counterparts, 
since they are expected to have lower rates of relaxation-driven stellar evaporation
\citep{McLaughlinFall07, Chandaretal07}.

\begin{figure}
\begin{center}
\psfig{figure=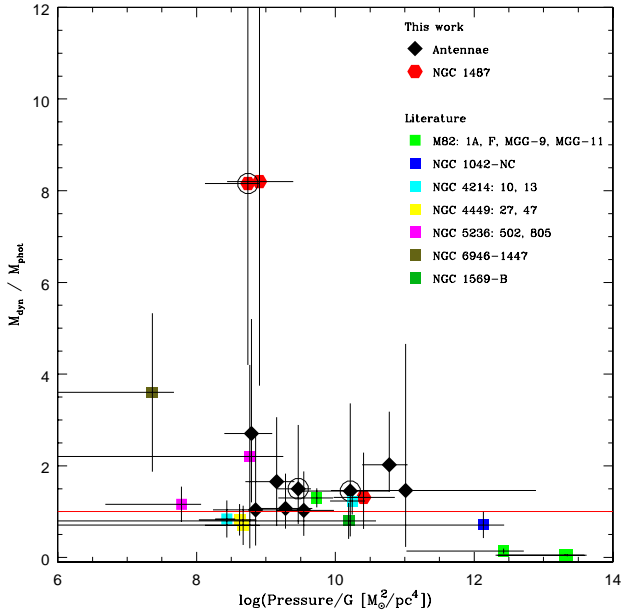,width=8.8cm}
\caption{The ratio of dynamical mass over photometric mass vs. pressure (taken as the log on the X-axis). 
Symbols, references and interpretation (except now for pressure, rather than
density) are the same as in Fig. \ref{massratiovsdensity}. \label{massratiovspressure}}
\end{center}
\end{figure}

A similar picture emerges if we consider pressure instead of density:
Following \citet{Elmegreen99}, we used $P \propto G M_{ph}^2 / r_{hp}^4$  
as an estimate for the pressure in the
ambient medium during cluster formation. Then SFE is expected to also
scale with this parameter \citep{Elmegreen99}. Indeed, as shown
in Figure~\ref{massratiovspressure}, the clusters which may be disolving are
amongst those with the lowest pressures in both our sample and the
sample taken from the literature.

Whereas low pressure/density is not a unique identifier for dissolving
cluster (since, as shown, there exist also low-density/pressure clusters with
no sign of expansion), about 50\% of the clusters in our plots below
a density/pressure limit (10$^{9.5}$ and 10$^9$, respectively) are 
dissolution candidates. They constitute only 20-25\% of the 
whole sample.

In our future work, we will compare dynamical and photometric mass estimates
of a larger sample, which would provide a more robust estimate of the fraction
of clusters which appear as single entities, but are unbound at an age of $\approx$10 Myr.

\section{Acknowledgements}

We are grateful to N. Bastian for making their modelling results available in electronic form, so
we could include them in our Fig. \ref{Clustersurvival}. We also wish to thank the anonymous referee
for constructive comments.


\begin{thebibliography}{}

\bibitem[Ascenso(2008)]{Ascenso2008} Ascenso J., to appear in the proceedings of the meeting ``Young massive star clusters - Initial conditions and environment'', E. Perez, R. de Grijs, R. M. Gonzalez Delgado, eds., Granada (Spain), September 2007, Springer: Dordrecht

\bibitem[Bastian et al.(2005)]{Bastian05} Bastian N., Gieles M., Lamers H.J.G.L.M., 
Scheepmaker R.A., de Grijs R., 2005, A\&A, 431, 905

\bibitem[Bastian et al.(2006)]{Bastian06} Bastian N., Saglia R.P., Goudfrooij P., 
Kissler-Patig M., Maraston C., Schweizer F., Zoccali M., 2006, a\&A, 448, 881

\bibitem[Baumgardt \& Kroupa(2007)]{BaumgardtKroupa2007} Baumgardt H., and Kroupa P., 2007, MNRAS, 380, 1589

\bibitem[Binney \& Tremaine(1987)]{binneyandtremaine87} Binney J., \& Tremaine S., 1987, "Galactic Dynamics",
Princeton Series in Astrophysics, (Princeton University Press; Princeton, NJ)

\bibitem[Boily \& Kroupa(2003a)]{BoilyKroupa03a} Boily C.M. \& Kroupa P., 2003a, MNRAS, 338, 665

\bibitem[Boily \& Kroupa(2003b)]{BoilyKroupa03b} Boily C.M. \& Kroupa P., 2003b, MNRAS, 338, 673

\bibitem[B\"oker et al.(2004a)]{Boeker04a} B\"oker T., Sarzi M., McLaughlin D.E., van der Marel R.P., 
Rix H.-W., Ho L.C., Shields J.C., 2004a, AJ, 127, 105
 
\bibitem[B\"oker et al.(2004b)]{Boeker04b} B\"oker T., Walcher C.J., Rix H.-W., H\"aring N., Schinnerer E., 
Sarzi M., van der Marel R.P., Ho L.C., Shields J.C., Lisenfeld U., Laine S., 2004b,
Proceedings of the conference ``The Formation and Evolution of Massive Young Star Clusters'',
ASP Conference Series, Vol. 322. Edited by H.J.G.L.M. Lamers, L.J. Smith, and A. Nota. San Francisco: 
Astronomical Society of the Pacific, p.39

\bibitem[Chandar, Fall \& McLaughlin(2007)]{Chandaretal07} Chandar R., Fall S.M., McLaughlin D.E., 2007,
ApJL, 668, 119

\bibitem[Chandar, Whitmore \& Lee(2004)]{ChandarWhitmoreLee04} Chandar R., Whitmore B.C, and Lee, 2004, ApJ, 611, 220
 
\bibitem[Chernoff \& Weinberg(1990)]{CW90} Chernoff D.F. \& Weinberg D.M., 1990, ApJ, 351, 121

\bibitem[Cohn(1979)]{cohn79} Cohn H., 1979, ApJ, 234, 1036

\bibitem[de Grijs, Wilkinson \& Tadhunter(2005)]{deGrijsetal05} de Grijs R., Wilkinson M.I., Tadhunter C.N.,
2005, MNRAS, 361, 311

\bibitem[De Marchi et al.(2004)]{DeMarchi04} De Marchi G., Sirianni M., Gilliland R., Bohlin R., 
Pavlovsky C., Jee M., Mack J., van der Marel R., and Boffi F., 2004, {\it Instrument Science Report ACS 2004-8}

\bibitem[Elmegreen et al.(2000)]{Elmegreen99} Elmegreen B.G., Efremov Y., Pudritz R.E., \& Zinnecker H., 2000,
Proceedings of the conference ``Protostars and Planets IV'', Tucson: University of Arizona Press; eds Mannings, V., 
Boss, A.P., Russell, S. S, p.179

\bibitem[Fall \& Zhang(2001)]{FallZhang01} Fall S.M., \& Zhang Q., 2001, ApJ, 561, 751

\bibitem[Fall(2004)]{fall04}Fall S. M. 2004, astro-ph/0405064

\bibitem[Fall, Chandar \& Whitmore(2005)]{Fall05} Fall S.M., Chandar R., \& Whitmore B.C., 2005, ApJ, 631, L133 

\bibitem[Fleck et al.(2006)]{Fleck06} Fleck J.-J., Boily C.M., Lan\c{c}on A, \& Deiters S., 2006, MNRAS, 369, 1392

\bibitem[Goodwin and Bastian(2006)]{GoodwinBastian06} Goodwin S.P., \& Bastian N., 2006, MNRAS 373, 752

\bibitem[Harris(1991)]{Harris91} Harris W.E., 1991, ARA\&A, 29, 543

\bibitem[Harris(1996)]{Harris96} Harris W.E., 1996, 1996, AJ, 112, 1487

\bibitem[Hills(1980)]{Hills80} Hills J.G., 1980, ApJ, 235, 986

\bibitem[Ho \& Filippenko(1996a)]{HF96a} Ho, L. \& Filippenko, A. 1996,
ApJ, 466, L83

\bibitem[Ho \& Filippenko(1996b)]{HF96} Ho, L. \& Filippenko, A. 1996,
ApJ, 472, 600

\bibitem[Holtzman et al.(1992)]{Hetal92} Holtzman J.A, et al., 1992, AJ, 103, 691

\bibitem[Jordan et al.(2006)]{Jordanetal06} Jord\'{a}n A., McLaughlin D.E., C\^{o}t\'{e} P., Ferrarese L., 
Peng E.W., Blakeslee J.P., Mei S., Villegas D., Merritt D., Tonry J.L., West M.J., 2006, ApJL, 651, 25 

\bibitem[Kouwenhoven \& de Grijs(2007)]{Kouwenhoven07} Kouwenhoven M.B.N., \& de Grijs R., 2007, astro-ph 0712.1748

\bibitem[King(1962)]{King62} King I.R., 1962, AJ, 67, 471

\bibitem[Krist(1995)]{Krist95} Krist, J.  1995, in ASP Conf.  Ser.  77,
Astronomical Data Analysis, Software, and Systems IV, eds. R. Shaw, H.
E. Payne, \& J. E. Hayes (San Francisco: ASP), 349

\bibitem[Lada et al.(1984)]{Lada84} Lada C.J., Margulis M., \& Dearborn D., 1984, 
ApJ, 285, 141

\bibitem[Lada \& Lada(2003)]{LadaLada} Lada C.J. \& Lada E.A., 2003, ARA\&A, 41, 57

\bibitem[Larsen(1999)]{Larsen99} Larsen S.S., 1999, A\&AS, 139, 393

\bibitem[Larsen(2001)]{Larsen01} Larsen S.S., 2001, AJ, 122, 1782

\bibitem[Larsen(2003)]{LarsenManual} Larsen S.S., 2003, ishape manual, distributed with
baolab-0.93.6 data reduction package

\bibitem[Larsen(2004)]{Larsen04} Larsen S.S., 2004, A\&A, 416, 537

\bibitem[Larsen \& Brodie(2000)]{LarsenBrodie00} Larsen S.S., Brodie J.P, 2000, AJ, 120, 2938

\bibitem[Larsen \& Richtler(2004)]{LarsenRichtler04} Larsen S.S. \&
Richtler T, 2004, A\&A, 427, 495

\bibitem[Larsen et al.(2004)]{Larsenetal04} Larsen S.S., Brodie
J.P, \& Hunter D.A., 2004, AJ, 128, 2295

\bibitem[Larsen et al.(2006)]{Larsenetal06} Larsen S.S., Brodie
J.P, \& Hunter D.A., 2006, AJ, 131, 2362

\bibitem[Larsen et al.(2008)]{Larsenetal08} Larsen S.S., Origlia L.,
Brodie J.P, Gallagher J.S., 2008, MNRAS, 383, 263


\bibitem[Larson(1999)]{Larson99} Larson R.B., 1999, in Proceedings...

\bibitem[Lee, Chandar \& Whitmore(2005)]{Leeetal05} Lee M.G., Chandar R., and Whitmore B.C., 
2005, 130, 2128

\bibitem[Lee \& Lee(2005)]{LeeLee} Lee H.J, \& Lee M.G., 2005, JKAS, 38, 345

\bibitem[Leitherer et al.(1999)]{Letal99} Leitherer, C., et al.  1999, ApJS, 123, 3

\bibitem[Maraston(2005)]{Maraston05} Maraston C., 2005, MNRAS, 362, 799

\bibitem[McCrady, Gilbert and Graham(2003)]{MGG03} McCrady N., Gilbert A.M., and Graham J.R., 2003, ApJ, 596, 240

\bibitem[McCrady, Graham and Vacca(2005)]{McCrady05} McCrady N., Graham J.R., and Vacca W.D., 2005, ApJ, 621, 278

\bibitem[McCrady \& Graham(2007)]{McCrady07} McCrady N., and Graham J.R., 2007, ApJ, 663, 844

\bibitem[McLaughlin \& Fall(2007)]{McLaughlinFall07}  McLaughlin D.E., and Fall S.M., 2007, astro-ph/0704.0080

\bibitem[Mengel et al.(2001)]{Mengel01} Mengel, S., Lehnert, M.D.,
Thatte, N., Tacconi-Garman, L. E., \& Genzel,~R. 2001, ApJ, 550, 280

\bibitem[Mengel et al.(2002)]{Mengel02} Mengel S., Lehnert M.D., Thatte
N., \& Genzel R., 2002, A\&A, 383, 137

\bibitem[Mengel et al.(2003)]{Mengel03} Mengel S., Lehnert M.D., Thatte N., \& Genzel R., 2003, SPIE, 4834, 45

\bibitem[Mengel et al.(2005)]{Mengel05} Mengel S., Lehnert M.D., Thatte N., \& Genzel R., 2005,
A\&A, 443, 41

\bibitem[Mengel and Tacconi-Garman(2007)]{MengelTacconiGarman07} Mengel S. \& Tacconi-Garman L.E., 2007, 
astro-ph/0701415

\bibitem[Meurer et al.(1995)]{Meurer95} Meurer G.R., Heckman T.M., Leitherer C., Kinney A.,
Robert C.,  \& Garnett D.R., 1995, AJ, 110, 2665

\bibitem[Peng et al.(2006)]{Peng06} Peng E.W., C\^{o}t\'{e} P., Jord\'{a}n A., Blakeslee J.P., 
Ferrarese L., Mei S., West M.J., Merritt D., Milosavljevi\'{c} M., Tonry J.L., 2006, ApJ, 639, 838

\bibitem[Persson et al.(1998)]{Persson98} Persson S.E., Murphy D.C., Krzeminski W., 
Roth M., Rieke M. J., 1998, AJ, 116, 2475

\bibitem[Saviane et al.(2008)]{Savianeetal08} Saviane I., Momany Y., Da Costa G.S.,
Rich R.M., Hibbard J., 2008, astro-ph/0802.1045

\bibitem[Sirianni et al.(2005)]{Sirianni05} Sirianni M., Jee M.J., Ben\'itez N., Blakeslee J.P.,
Martel A.R., Meurer G., Clampin M., De Marchi G., Ford H.C., Gilliland R., Hartig G.F., 
Illingworth G.D., Mack J., and McCann W.J., 2005, PASP, 117, 1049

\bibitem[Smith et al.(2006)]{Smith06} Smith L.J., Westmoquette M.S., Gallagher III J.S., O'Connell R.W.,
Rosario D.J., and de Grijs R., 2006, MNRAS, 370, 513

\bibitem[Spitzer(1978)]{spitzer78} Spitzer L. Jr., 1978, {\em  Physical Processes in the
Interstellar Medium}, Wiley-Interscience: New York

\bibitem[Spitzer(1987)]{spitzer87} Spitzer L. Jr., 1987, {\em Dynamical Evolution of Globular
Clusters}, Princeton: Princeton Univ. Press, 1987

\bibitem[Sternberg(1998)]{Sternberg98} Sternberg, A., 1998, ApJ, 506, 721

\bibitem[Takahashi \& Portegies Zwart(2000)]{TPZ00}Takahashi, K., \&
Portegies Zwart, S.  F.  2000, ApJ, 535, 759

\bibitem[Trancho et al.(2007)]{Trancho07}Trancho G., Bastian N., Miller B.W., Schweizer F., 
2007, ApJ, 664, 284

\bibitem[van den Bergh(1991)]{vandenBergh91}van den Bergh S., 1991, ApJ, 369, 1

\bibitem[Vazquez \& Leitherer(2005)]{VazquezLeitherer05}Vazquez G.A.  \&
Leitherer C., 2005, ApJ, 621, 695

\bibitem[Waters et al.(2006)]{Watersetal06} Waters Ch.Z., Zepf S.E., Lauer T.R., Baltz E.A.; Silk J.,
2006, ApJ, 650, 885

\bibitem[Whitmore et al.(1993)]{Wetal93} Whitmore B.C., Schweizer F., Leitherer C., Borne K., \&
Robert C., 1993, AJ, 106, 1354

\bibitem[Whitmore \& Schweizer(1995)]{WS95} Whitmore, B. \& Schweizer,
F. 1995, AJ, 109, 960

\bibitem[Whitmore et al.(1997)]{Wetal97} Whitmore B.C, Miller B.W., Schweizer F.,  \&
 Fall S.M., 1997,  AJ, 114, 2381

\bibitem[Whitmore et al.(1999)]{W99}Whitmore B.C., Zhang, Q., Leitherer, C.
Fall, S. M., Schweizer, F., \& Miller, B. W. 1999, AJ, 118, 1551

\bibitem[Whitmore et al.(2000)]{Whitmore2000}Whitmore B.C., 2000, astro-ph/0012546

\bibitem[Whitmore(2004)]{W04}Whitmore B.C., 2004, in The Formation and Evolution of Massive
Young Star Clusters, ASP Conference Series, Vol. 322, Ed. H.J.G.L.M. Lamers, L.J. Smith, and A. Nota. San Francisco:
Astronomical Society of the Pacific, p. 419

\bibitem[Whitmore(2007)]{Whitmore07} Whitmore B.C., Chandar R., and Fall S.M., 2007, AJ, 133, 1067

\bibitem[Wilson et al.(2001)]{Wilson01} Wilson, C.  D., Scoville, N.,
Madden, S.  C., \& Charmandaris, V.  2000, ApJ, 542, 120

\bibitem[Zhang \& Fall(1999)]{ZhangFall99} Zhang, Q. \&  Fall, S. M. 1999,
ApJ, 527, 81

\bibitem[Zepf and Ashman(1999)]{Zepf99} Zepf S.E, \& Ashman K.M., 1999, AJ, 118, 752

\end{thebibliography}
\end{document}